\documentclass[showpacs,amsmath,amssymb,floatfix,prl,twocolumn]{revtex4}
\usepackage{graphicx,amsmath}

\begin{document} 

\title {PageRank model of opinion formation on social networks}

\author{Vivek Kandiah}
\affiliation{\mbox{Laboratoire de Physique Th\'eorique du CNRS, IRSAMC, 
Universit\'e de Toulouse, UPS, F-31062 Toulouse, France}}
\author{Dima L. Shepelyansky}
\affiliation{\mbox{Laboratoire de Physique Th\'eorique du CNRS, IRSAMC, 
Universit\'e de Toulouse, UPS, F-31062 Toulouse, France}}

\date{April 17, 2012}


\begin{abstract}
We propose the PageRank model of opinion formation
and investigate its rich properties on real directed networks
of Universities of Cambridge and Oxford, LiveJournal and Twitter.
In this model the opinion formation of linked electors is
weighted with their PageRank probability.
We find that the society elite, corresponding to the top PageRank
nodes, can impose its opinion to a significant fraction of the
society. However, for a homogeneous distribution of 
two opinions   there  exists a 
bistability range of opinions which 
depends on a conformist parameter characterizing 
the opinion formation. We find that LiveJournal and Twitter
networks have a stronger tendency to a totalitar opinion formation.
We also analyze the Sznajd model generalized for scale-free networks
with the weighted PageRank vote of electors. 
\end{abstract}

\maketitle

\section{I. Introduction}

To understand the nature and origins of mass opinion formation
is the outstanding challenge of democratic societies \cite{zaller}. 
In a last few years an enormous development of such social networks
as LiveJournal \cite{livejournal}, 
Facebook \cite{facebook}, Twitter \cite{twitter} and
VKONTAKTE \cite{vkontakte}, with up to hundreds of millions of users,
demonstrated the growing
influence of these networks on social and political life.
The small-world scale-free structure of the social networks
(see e.g. \cite{dorogovtsev,caldarelli}),
combined with their rapid communication facilities,  
leads to a very fast information propagation over 
networks of electors, consumers, citizens
making them very active on instantaneous social events.
This puts forward a request for new theoretical models
which would allow to understand 
the opinion formation process in the modern society
of XXI century.

The important steps in the analysis of opinion formation 
have been done with the development of 
various voter models described
in a great detail in \cite{galam0},
\cite{liggett},\cite{galamepl},\cite{watts2007},\cite{galam},
\cite{fortunatormp},\cite{krapivskybook},\cite{schmittmann}.
This research field became known as sociophysics 
\cite{galam0},\cite{galamepl},\cite{galam}.
In this work we introduce several new aspects which 
take into account the generic features of the social networks.
At first, we analyze the opinion formation
on real directed networks taken from the Academic Web Link Database 
of British universities networks \cite{ukweb},
LiveJournal database \cite{benczur} and Twitter dataset \cite{twitters}. 
This allows us to
incorporate the correct scale-free network structure
instead of unrealistic  regular lattice networks,
often considered in voter models \cite{fortunatormp,krapivskybook}.
At second, we assume that the opinion
at a given node is formed by the opinions
of its linked neighbors weighted with the PageRank probability
of these network nodes. We think that this step represents
the reality of social networks: all of network nodes are characterized
by the PageRank  vector which gives a probability to find a random surfer on
a given node as described in \cite{brin,meyerbook}.
This vector gives a steady-state probability distribution on the network
which provides a natural ranking of node importance,
or elector or society member importance.
In a certain sense the top nodes of PageRank correspond to a political elite
of the social network which opinion influences the opinions
of other members of the society \cite{zaller}. 
Thus the proposed PageRank Opinion Formation (PROF) 
model takes into account the situation in which
an opinion of an influential friend from high ranks of the society
counts more than an opinion  of a friend from lower society level.
We argue that the PageRank probability is the most natural
form of ranking of society members. Indeed, the efficiency of PageRank
rating is demonstrated 
for various types of scale-free networks including
the World Wide Web (WWW) \cite{brin,meyerbook},
{\it Physical Review} citation network \cite{rednerphystod,fortunatopre},
scientific journal rating \cite{eigenfactor},
ranking of tennis players \cite{tennis},
Wikipedia articles   \cite{wiki}, the world trade network
\cite{wtrade} and others. Due to the above argument we consider that the PROF model
captures the reality of social networks and below we present 
the analysis of its interesting properties.

The paper is composed as follows: the PROF model is described in Sec. II,
the numerical results on  its properties are presented in
Sec.III for British University networks. 
In Sec. IV we combine the PROF model with the
Sznajd model \cite{sznajd,fortunatormp} and 
study the properties of the PROF-Sznajd model.
In Sec.V we analyze the models discussed in previous Sections
on an example of  large social network of  the LiveJournal \cite{benczur}.
The results for the Twitter dataset \cite{twitters}
are presented in Sec.VI.
The discussion of the results is presented in Sec.VII.

\section{II. PageRank Opinion Formation (PROF) model description}
 
The PROF model is defined in the following way. 
In agreement with the standard PageRank algorithm \cite{meyerbook}
we determine the
PageRank probability  $P_i$ for each node $i$ and arrange all $N$ nodes
in a monotonic decreasing order of the probability.
In this way each node $i$ has a probability $P(K_i)$
and the PageRank index $K_i$ with the maximal
probability is at $K_i=1$ ($\sum_{i=1}^N P(K_i)=1$).
We use the usual damping factor value $\alpha=0.85$
to compute the PageRank vector of the Google matrix of the
network (see e.g. \cite{brin},\cite{meyerbook},\cite{frahm,2dmotor}).
In addition to that a network node
$i$ is characterized by an Ising spin variable $\sigma_i$
which can take values $+1$ or $-1$  coded also by red or blue
color respectively. The sign of a node $i$ is determined by its direct
neighbors $j$ which have the PageRank probabilities $P_j$.
For that we compute the sum  $\Sigma_i$ over 
all directly linked neighbors $j$ of node $i$:
\begin{eqnarray}
\label{eq1}
\Sigma_i=a\sum_j P^+_{j,in} + b\sum_j P^+_{j,out} \\ \nonumber
- a\sum_j P^-_{j,in} - b\sum_j P^-_{j,out} \;\;, \;\; a+b=1 \;\;  ,
\end{eqnarray}
\noindent
where $P_{j,in}$ and $P_{j,out}$ denote the PageRank probability
$P_j$  of a node $j$ pointing to node $i$ 
(incoming link) and a node $j$ to which node $i$ points to 
(outgoing link) respectively. Here, the two 
parameters $a$ and $b$ are used to tune the importance 
of incoming and outgoing links with the imposed relation
$a+b=1$ ($0 \leq a,b \leq 1$). The values $P^+$ and $P^-$
correspond to red and blue nodes respectively.
The value of spin $\sigma_i$
takes the value $1$ or $-1$ respectively for
$\Sigma_i>0$ or $\Sigma_i<0$. In a certain sense we can say that 
a large value of parameter $b$ corresponds to
a conformist society where an elector $i$ takes an 
opinion of other electors to which he points to
(nodes with many incoming links are 
in average at the top positions of PageRank).
On the opposite side  a large value of $a$ 
corresponds to a tenacious society
where an elector $i$ takes mainly an 
opinion of those electors  who point to him.
\begin{figure}[ht]
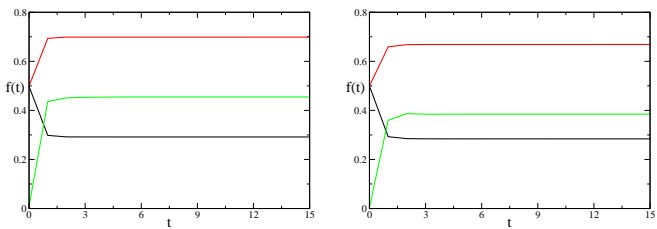

\vglue +0.2cm
\begin{center}
\includegraphics[width=0.23\textwidth]{fig1a}
{\hskip 0.2cm}
\includegraphics[width=0.23\textwidth]{fig1b}
\end{center}
\vglue -0.2cm
\caption{(Color online) Time evolution of opinion given
 by a fraction of  red nodes $f(t)$
as a function of number of iterations $t$.
The red  and black curves
(top and bottom curves at $t=15$ respectively)
show evolution for two different realizations of random
distribution of color with the same initial fraction
$f_i=0.5$ at $t=0$.
The green curve 
(middle curve at $t=15$)
shows dependence $f(t)$
for the initial state with $N_{top}$
all red nodes with top PageRank $K$ indexes
(highest $P(K_i)$ values, $1\leq K \leq N_{top}$).
The evolution is done at 
$a=b=0.5$ and temperature $T=0$. 
\emph{Left panel :} Cambridge network with  $N_{top}=2000$.
\emph{Right panel :} Oxford network with $N_{top}=1000$.
\label{fig1}}
\end{figure}

The condition (\ref{eq1}) on spin inversion
can be written via the effective Ising Hamiltonian $H$
of the whole system of interacting spins:
\begin{equation}
\label{eq2}
H=-\sum_{i,j} J_{ij} \sigma_i \sigma_j=
-\sum_i B_i \sigma_i=\sum_i \epsilon_i \; ,
\end{equation}   
where the spin-spin interaction $J_{ij}$
determines the local magnetic field $B_i$
on a given node $i$
\begin{eqnarray}
\label{eq3}
B_i=\sum_j (a P_{j,in} + b P_{j,out})  \sigma_j  \; ,
\end{eqnarray}
which gives the local spin energy $\epsilon_i=-B_i \sigma_i$.
According to (\ref{eq2}), (\ref{eq3}) the
interaction between a selected spin $i$ and 
its neighbors $j$  is given by the PageRank probability:
$J_{ij}=  a P_{j,in} + b P_{j,out}$.
Thus from a physical view point the whole system
can be viewed as a disordered ferromagnet \cite{galam,krapivskybook}.
In this way the condition (\ref{eq1})
corresponds to a local energy $\epsilon_i$  minimization
done at zero temperature. We note that such an analogy
with spin systems is well known for opinion formation
models on regular lattices 
\cite{galam},\cite{fortunatormp},\cite{krapivskybook}.
However, it should be noted that generally we have asymmetric
couplings $J_{ij} \neq J_{ji}$
that is unusual for physical problems
(see discussion in \cite{galam1}).
In view of this analogy it is possible to
introduce a finite temperature $T$
and then to make a probabilistic 
Metropolis type condition \cite{metropolis} for the spin $i$
inversion determined by a thermal probability 
$\rho_i = \exp(- \Delta \epsilon_i/T)$,
where $\Delta \epsilon_i$ is the energy difference 
between on-site energies $\epsilon_i$
with spin up and down. 
During the relaxation process each spin 
is tested on inversion condition 
that requires $N$ steps and then we do
$t$ iterations of such $N$ steps.
We discuss the results of the relaxation process 
at zero and finite temperatures $T$ in next Section.

\section{III. Numerical results for PROF model 
on university networks}

Here we present results for PROF model considered on the networks
of Cambridge and Oxford Universities in year 2006, taken from \cite{ukweb}.
The properties of PageRank distribution $P(K)$
for these networks have been analyzed in \cite{frahm,2dmotor}.
The total number of nodes $N$ and links $N_\ell$ are: 
$N=212710$, $N_\ell=2015265$
(Cambridge);  $N=200823$, $N_\ell=1831542$ (Oxford) \cite{2dmotor}.
Both networks are characterized by an algebraic decay of 
PageRank probability $P(K) \propto 1/K^\beta$ and approximately
usual exponent value $\beta \approx 0.9$,
 additional results on the scale-free properties 
of these networks are given in \cite{frahm,2dmotor}. We discuss 
usually the fraction of red nodes
since by definition all other nodes are blue.

\begin{figure}[ht]
\begin{center}
\includegraphics[width=0.23\textwidth]{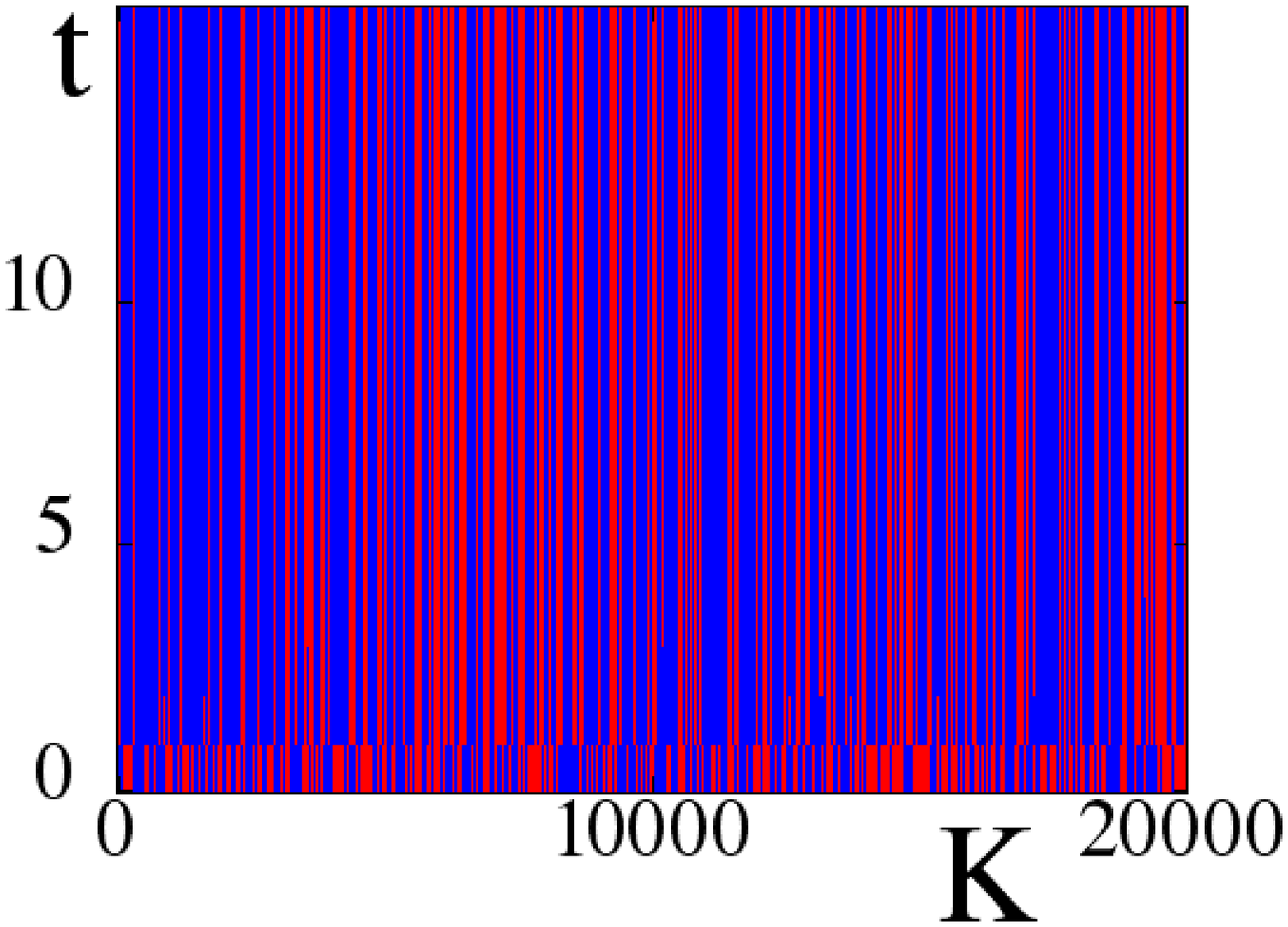}
\includegraphics[width=0.23\textwidth]{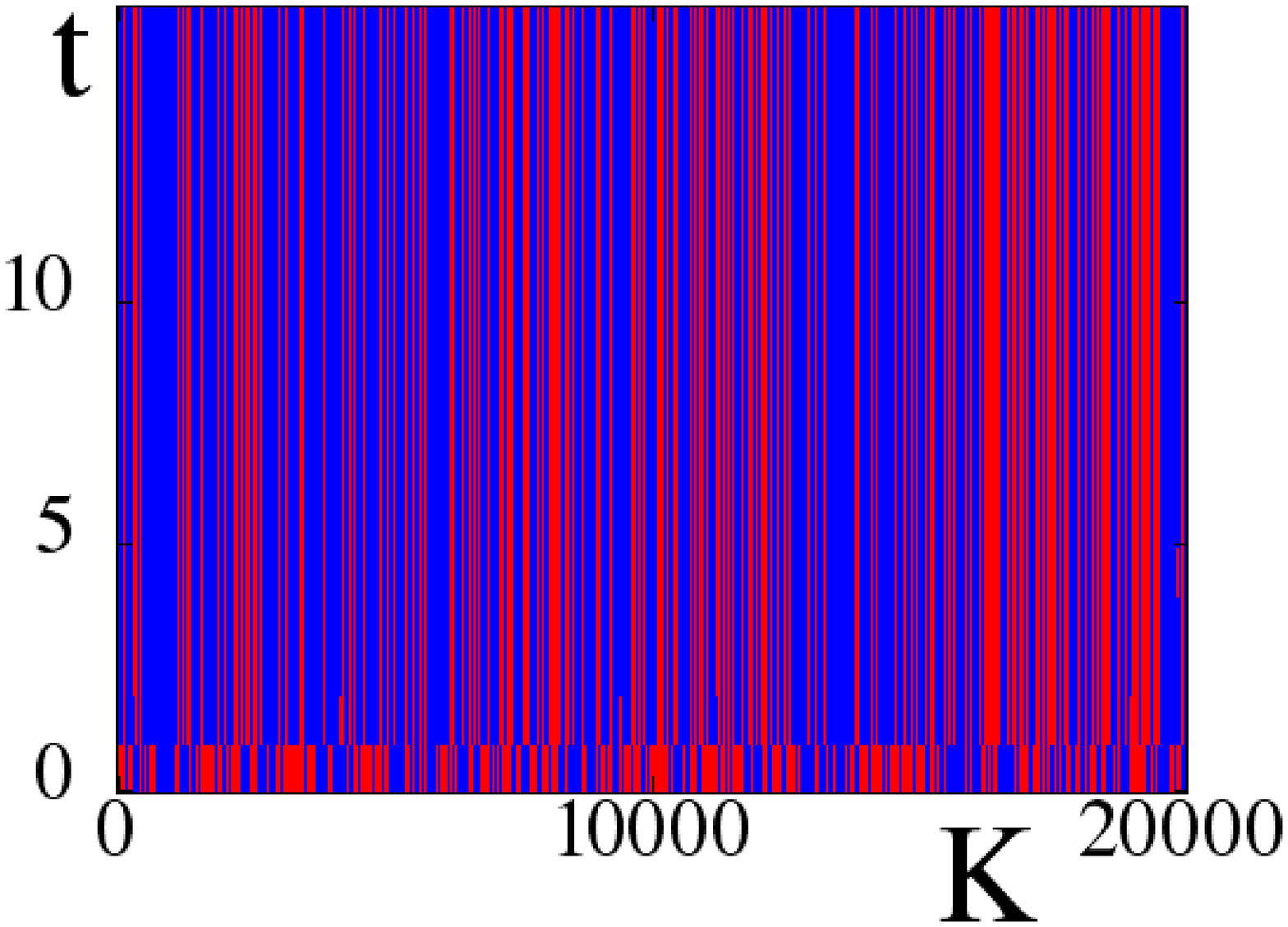}\\
\includegraphics[width=0.23\textwidth]{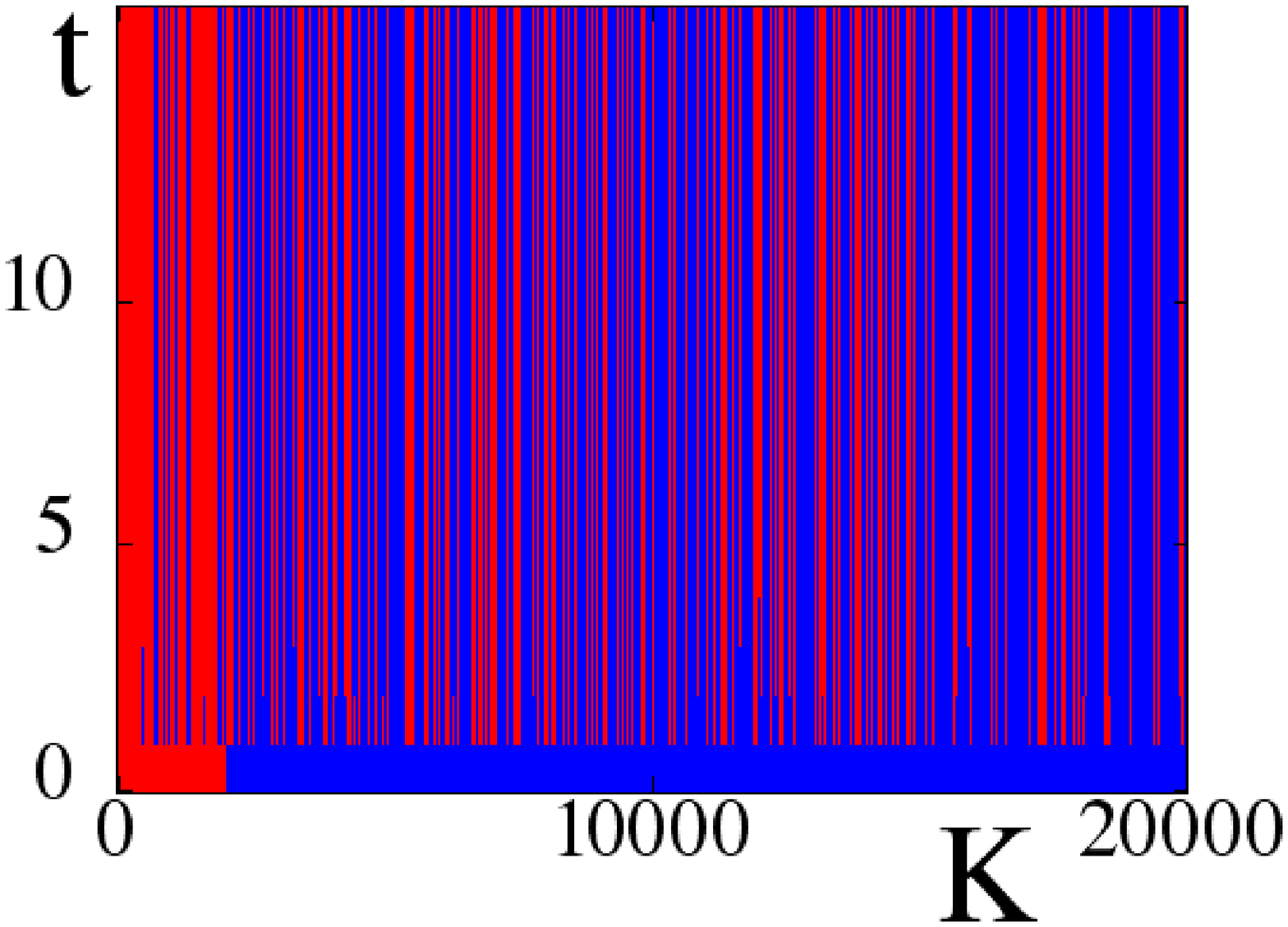}
\includegraphics[width=0.23\textwidth]{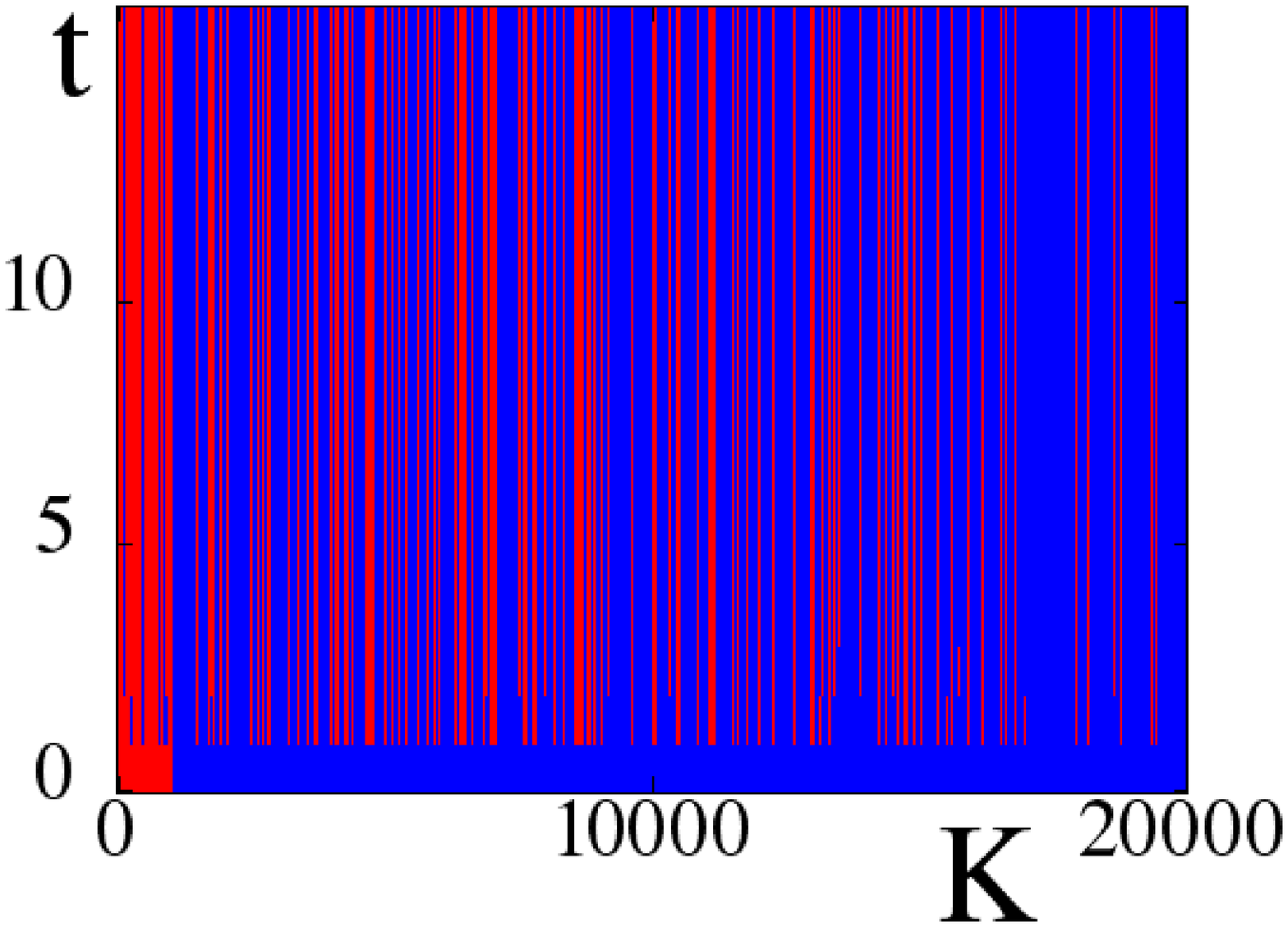}
\end{center}
\vglue -0.2cm
\caption{(Color online) Time evolution of opinion colors
(red/gray and blue/black)
for the parameters of Fig.~\ref{fig1}:
left/right column is for Cambridge/Oxford network.
The initial fraction of red  colors is
$f_i=0.5$ (top panel) and $N_{top}$ nodes
have red color for bottom panels with $N_{top}=2000$ and $1000$
for Cambridge and Oxford network  respectively.
Nodes are ordered by the PageRank index $K$
and color plot  shows  only $K \leq 20000$. 
\label{fig2}} 
\end{figure}

The typical examples of time evolution of
fraction of red nodes $f(t)$ with the number of 
time iterations $t$ are shown in Fig.~\ref{fig1}.
We see the presence of bistability in the opinion formation:
two random states with the same initial fraction of red nodes $f_i=f(t=0)$
evolve to two different final fractions of red nodes $f_f$.
The process gives an impression of convergence to a fixed
state approximately after $t_c \approx 10$ iterations.
A special check shows that all node colors
become fixed after this time $t_c$. The convergence time
to a fixed state is similar to those found for
opinion formation on regular lattices where
$t_c =O(1)$ \cite{fortunatormp,krapivskybook,rednerprl}.
The corresponding time evolution of colors is shown in Fig.~\ref{fig2}
for first 10\% of nodes ordered by the PageRank index $K$.

The results of Fig.~\ref{fig1} show that
for a random initial distribution of colors
we may have different final states with $\pm 0.2$
variation compared to the initial $f_i=0.5$.
However, if we consider that $N_{top}$ nodes with the top $K$ index values
(from $1$ to $N_{top}$) have the same opinion (e.g. red nodes)
then we find that  even a small fraction of the total number of nodes
$N$ (e.g. $N_{top}$ of about 0.5\% or 1\% of $N$) can 
impose its opinion for a significant 
fraction of nodes of about $f_f \approx 0.4$.
This shows that in the frame of PROF model the society elite,
corresponding to top $K$ nodes,
can significantly influence the opinion of the whole society
under the condition that the elite members have the fixed opinion
between themselves. 

We also considered the case when the red nodes
are placed on $N_{top} =2000$ top nodes of  CheiRank
index $K^*$. This ranking is characterized 
by the CheiRank probability $P^*(K^*)$ for a random surfer
moving in the inverted direction of links as described in \cite{wiki,2dmotor}.
In average $P^*(K^*)$ is proportional to the number of outgoing links.
However, in this case the top nodes with a small 
$f_i$ values are not able to impose their opinion
and the final fraction becomes blue. We attribute this to
the fact that the opinion condition (\ref{eq1}) is
determined by the PageRank probability $P(K)$
and that the correlations between CheiRank and PageRank
are not very strong (see discussion in \cite{wiki,2dmotor}).

\begin{figure}[ht]
\begin{center}
\includegraphics[width=0.23\textwidth]{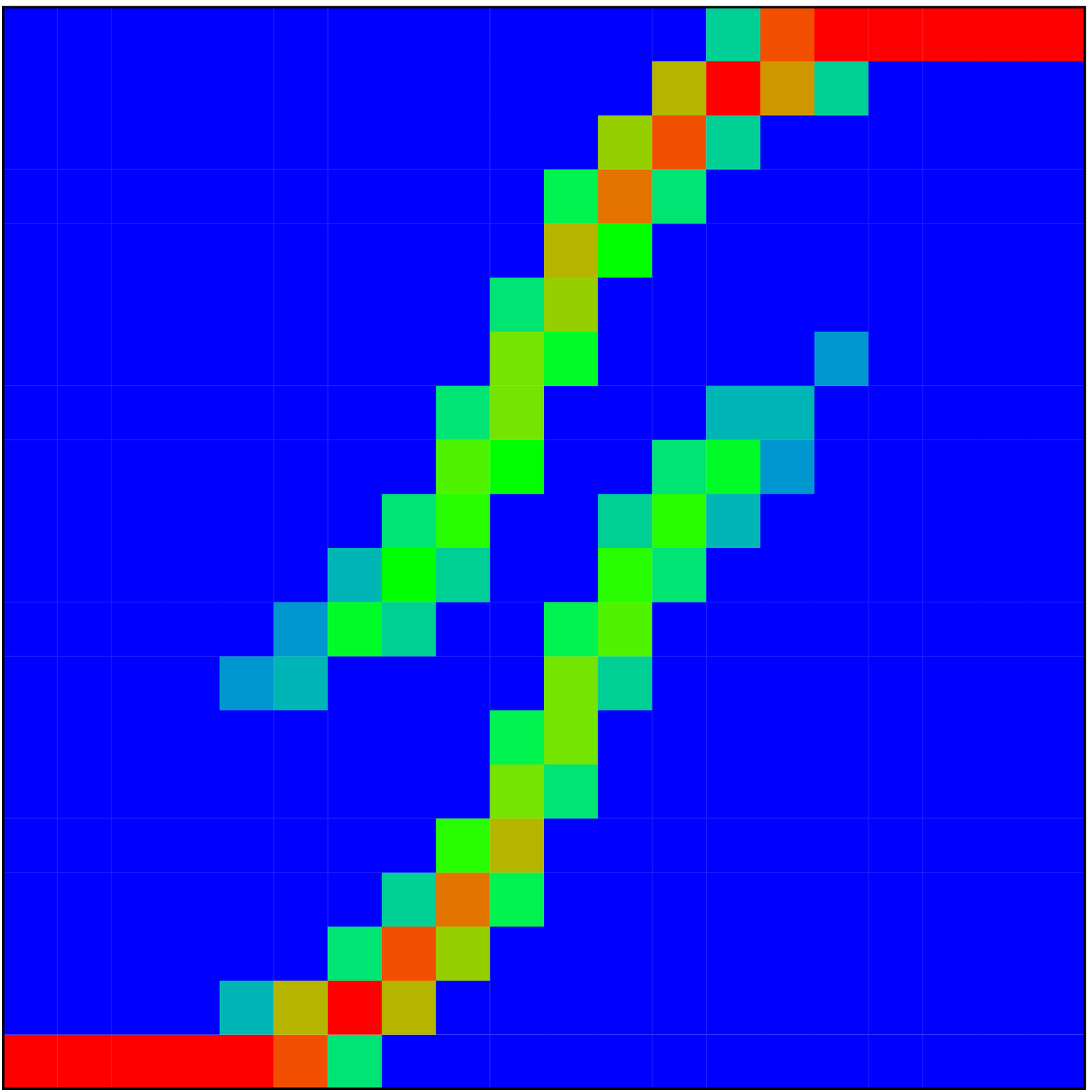}
\includegraphics[width=0.23\textwidth]{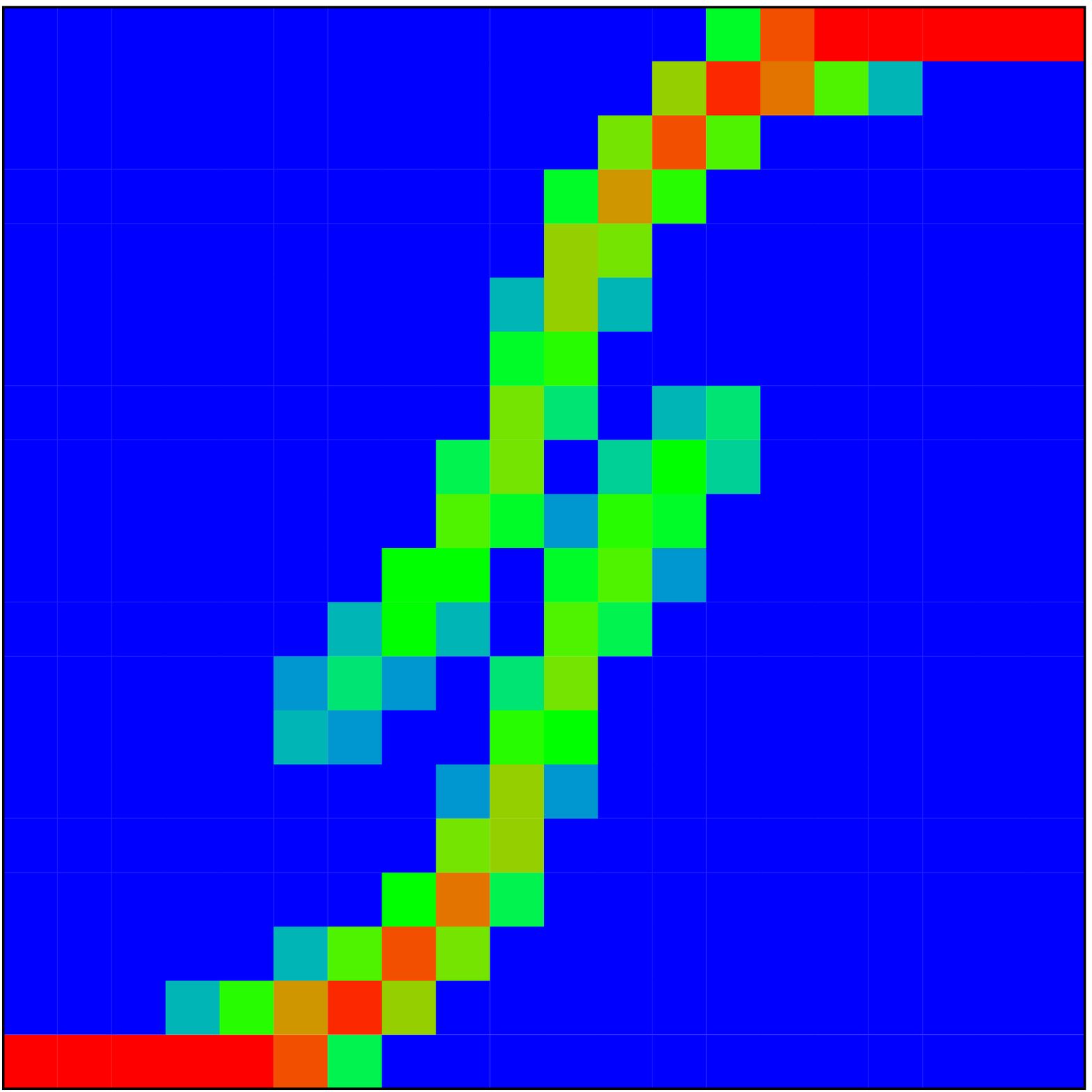}\\
\includegraphics[width=0.23\textwidth]{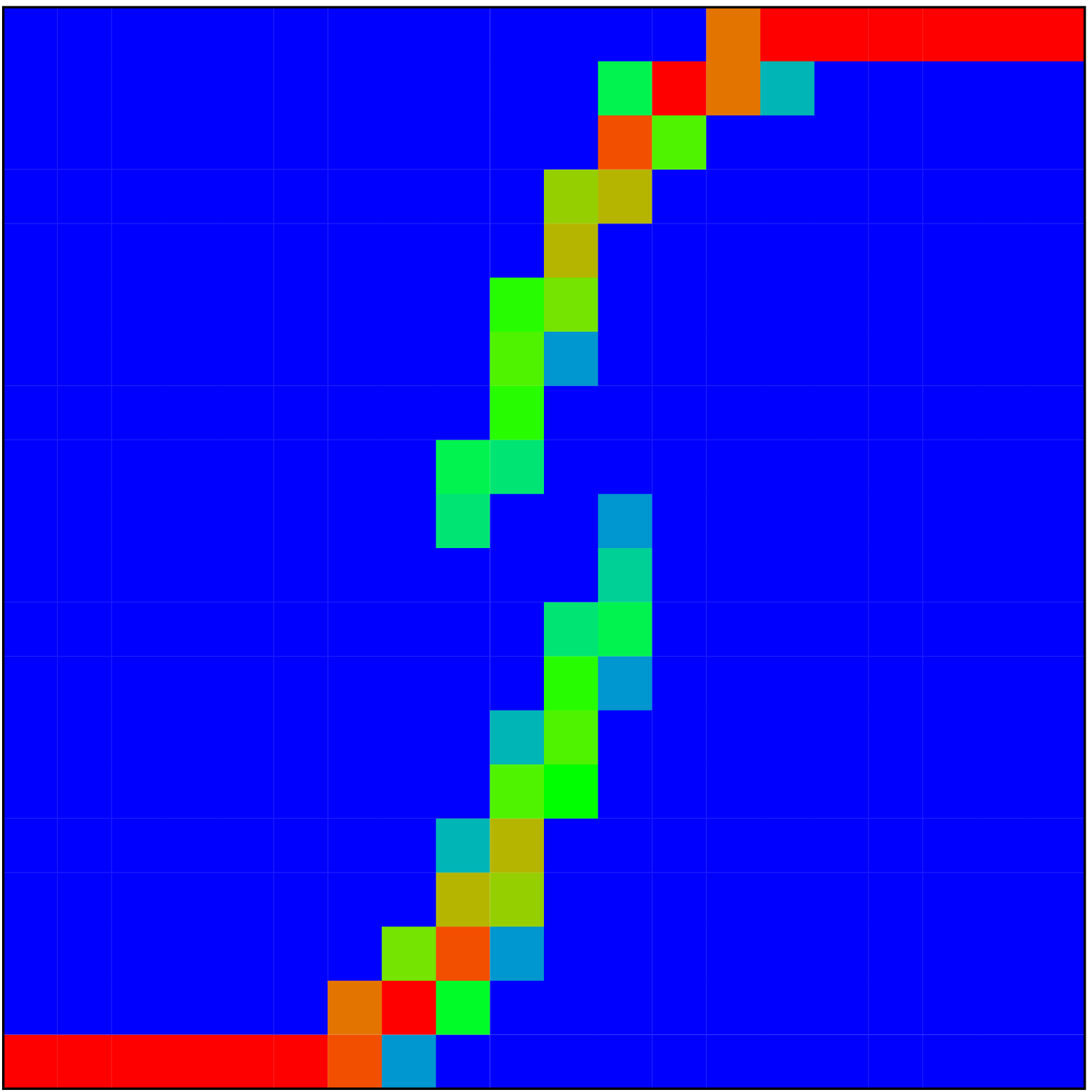}
\includegraphics[width=0.23\textwidth]{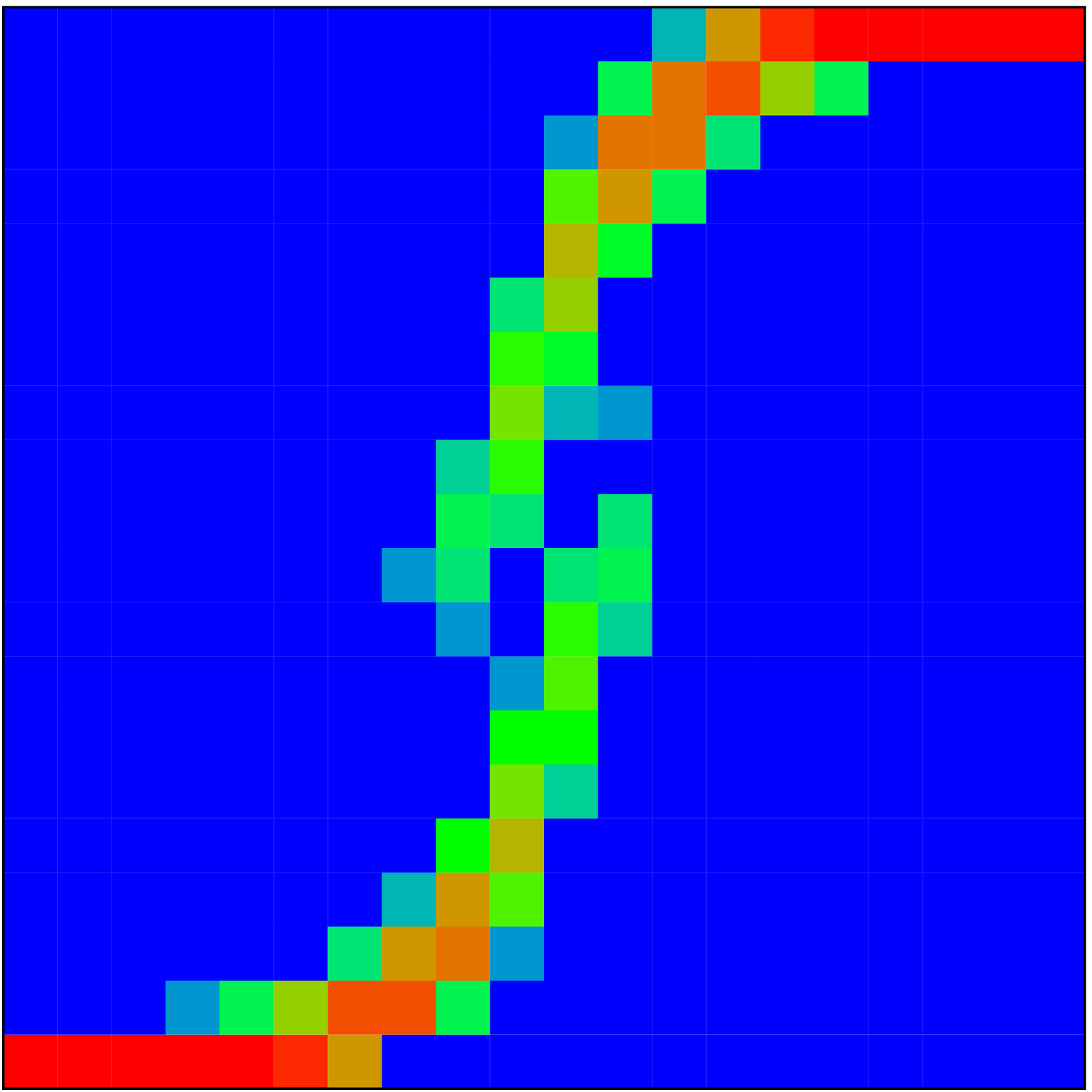}\\
\includegraphics[width=0.23\textwidth]{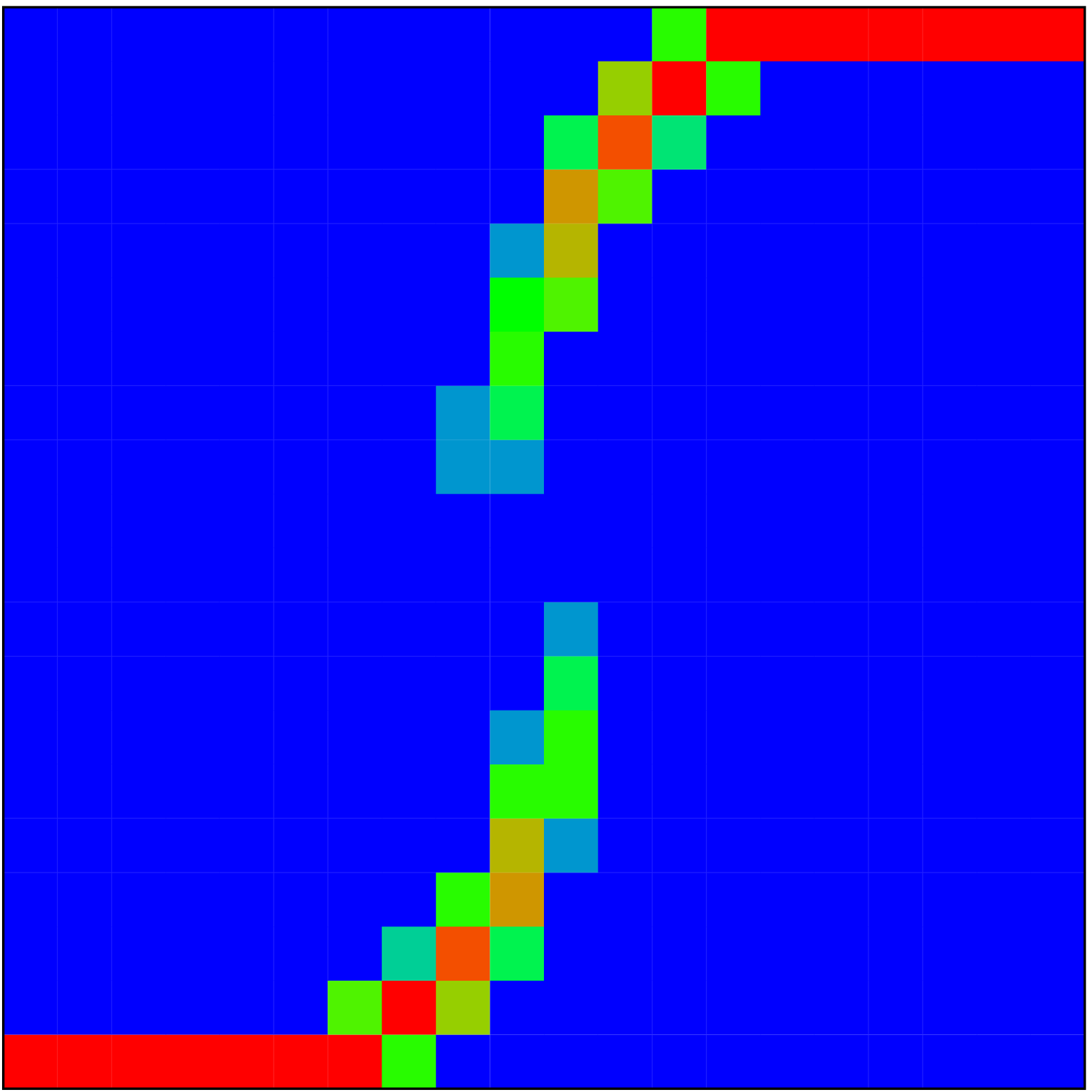}
\includegraphics[width=0.23\textwidth]{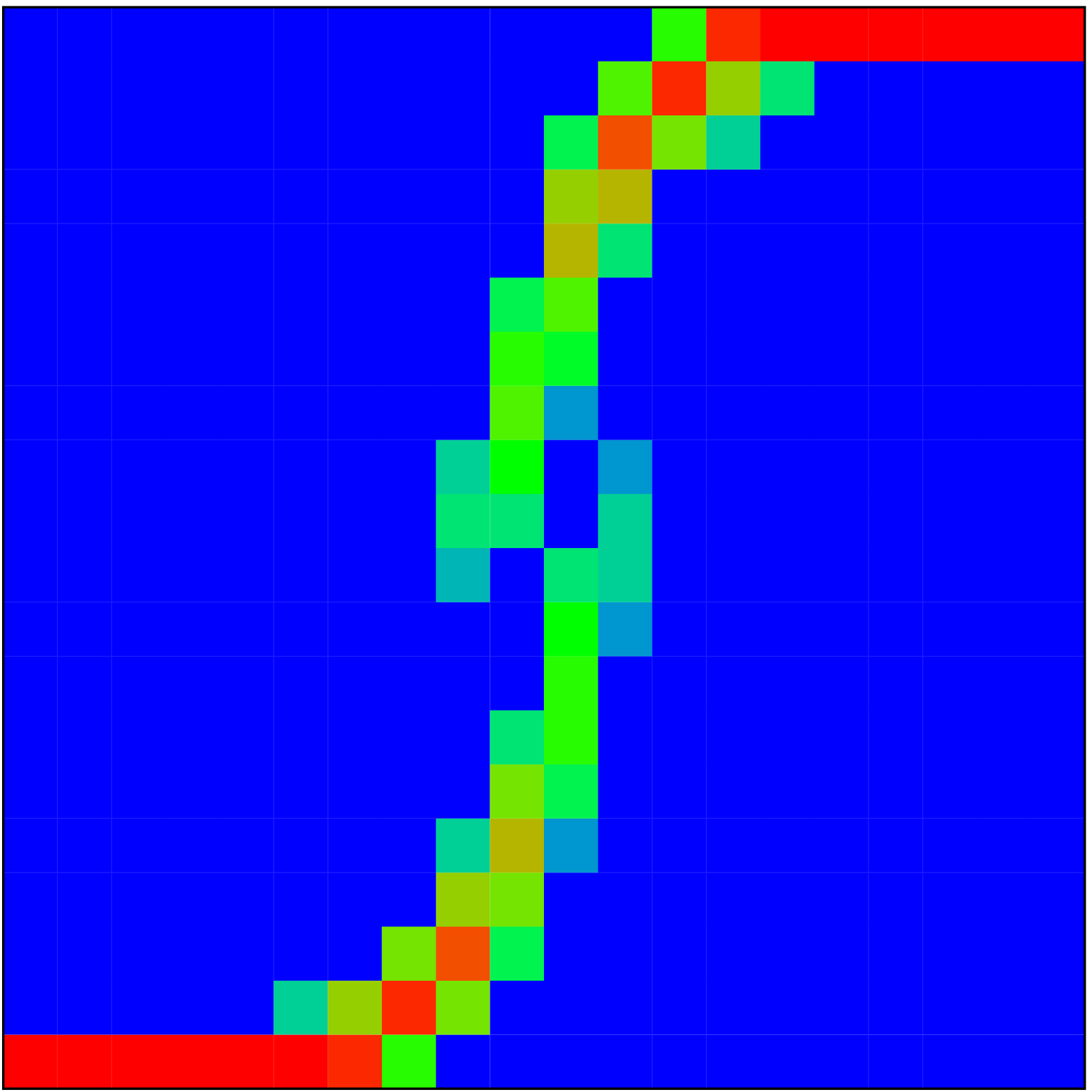}
\end{center}
\vglue -0.2cm
\caption{(Color online) Density plot of probability $W_f$ 
to find a final red fraction $f_f$, shown in $y-$axis,
in dependence on an initial red fraction $f_i$, shown in $x-$ axis;
data are shown inside the unit square $0\leq f_i,f_f \leq 1$.
The values of $W_f$ are defined as a relative number of
realizations found inside each of $20\times20$ cells
which cover  the whole unit square. Here $N_r=10^4$ realizations of
randomly distributed colors are used to obtained $W_f$ values;
for each realization the
time evolution is followed up the convergence time with up to 
$t=20$ iterations; here $T=0$.
{\it Left column:} Cambridge network;
{\it right column:} Oxford network;
here $a=0.1, 0.5, 0.9$ from top to bottom.
The probability $W_f$ is proportional to
color changing from zero (blue/black) to unity
(red/gray).
\label{fig3}} 
\end{figure}

To analyze how the final fraction of red nodes $f_f$
depends on its initial fraction $f_i$
we study the time evolution $f(t)$ for a large number $N_r$ of 
initial random realizations of colors following it up to the
convergent time for each realization. 
We find that the final red nodes are homogeneously
distributed in $K$. Thus there is no
specific preference for top society levels
for an initial random distribution.
The probability distribution
$W_f$ of final fractions $f_f$ is shown in Fig.~\ref{fig3}
as a function of initial fraction $f_i$ at three values of
parameter $a$. These results show two main features of the model:
a small fraction of red opinion is completely suppressed if
$f_i < f_{c}$ and its larger fraction 
dominates completely for $f_i>1-f_c$;
there is a bistability phase for the initial opinion range
$f_b \leq f_i \leq 1-f_b$.
Of course, there is a symmetry in respect to exchange
of red and blue colors. For small value $a=0.1$ we have $f_b \approx f_c$
with $f_c \approx 0.25$ while for large value $a=0.9$
we have $f_c \approx 35$, $f_b \approx 0.45$. 

\begin{figure}[ht]
\vglue +0.7cm
\begin{center}
\includegraphics[width=0.44\textwidth]{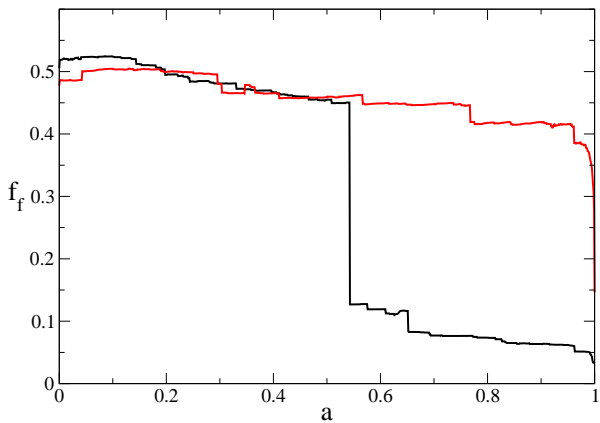}
\end{center}
\vglue -0.2cm
\caption{(Color online) Dependence of the final fraction of red nodes $f_f$
on the tenacious parameter $a$ (or conformist parameter $b=1-a$)
for initial red nodes in  $N_{top}=2000$ values of PageRank index
($1\leq K \leq N_{top}$); black and red(gray) curves
show data for Cambridge and Oxford networks; here $T=0$.
\label{fig4}}
\end{figure}

\begin{figure}[ht]
\begin{center}
\includegraphics[width=0.23\textwidth]{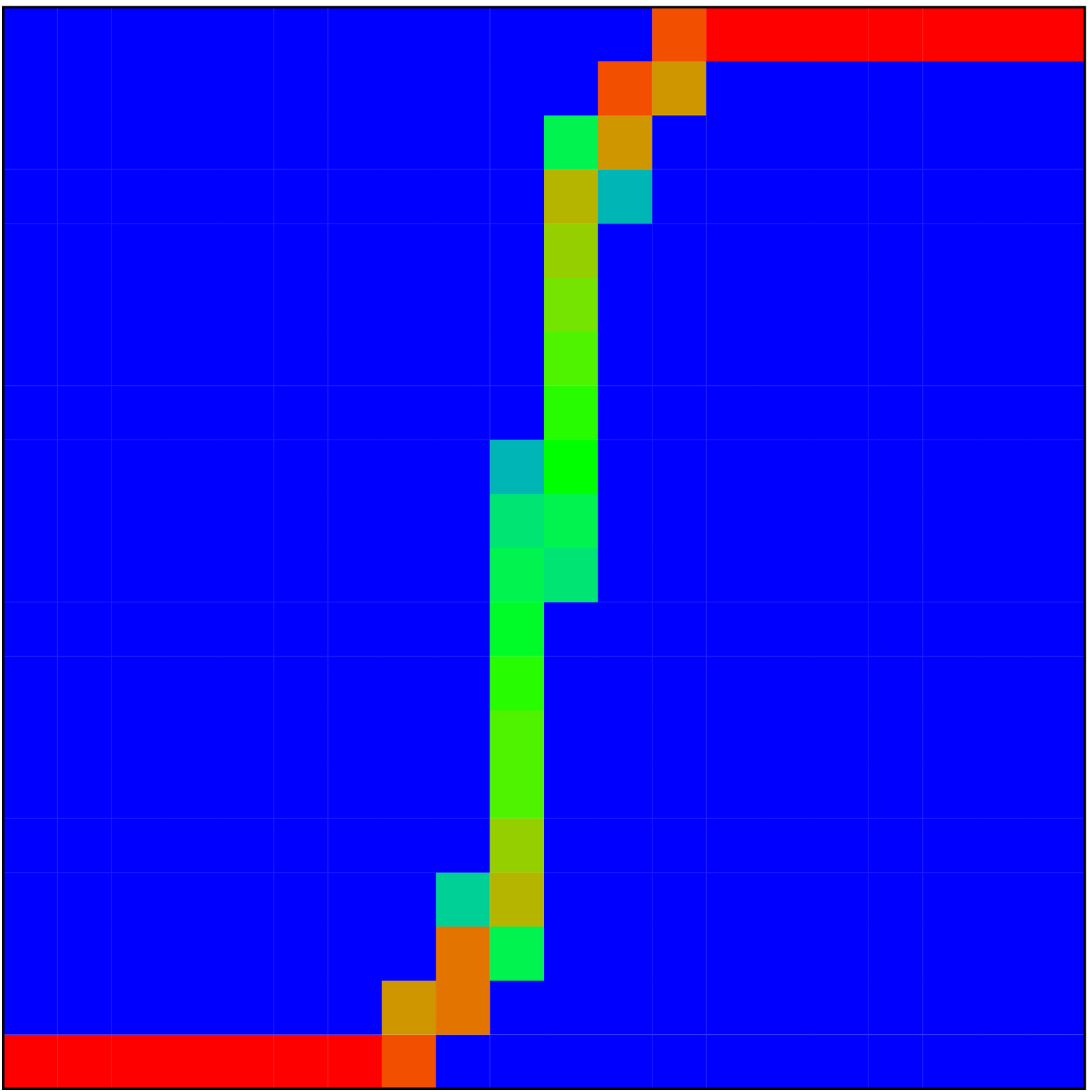}
\includegraphics[width=0.23\textwidth]{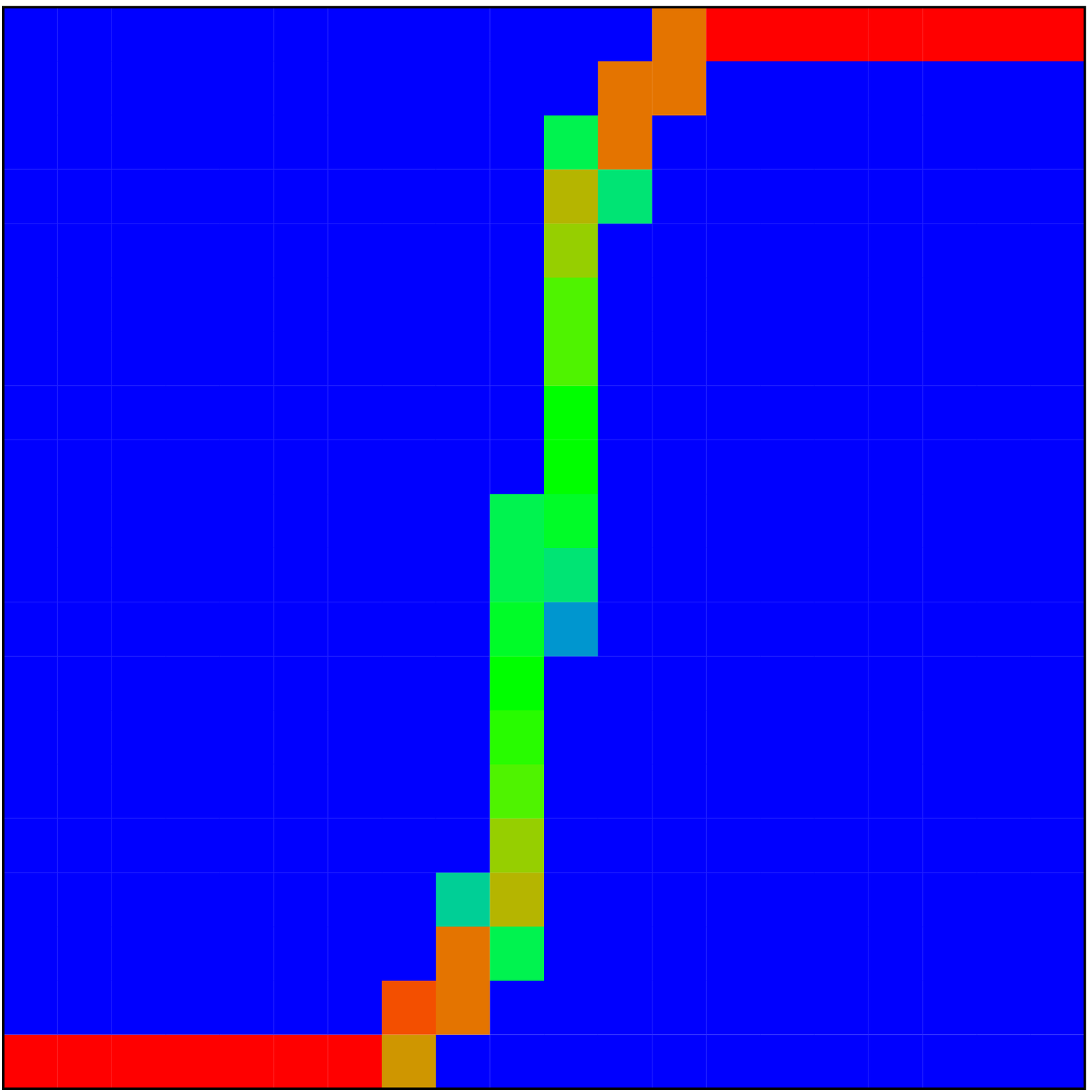}\\
\includegraphics[width=0.23\textwidth]{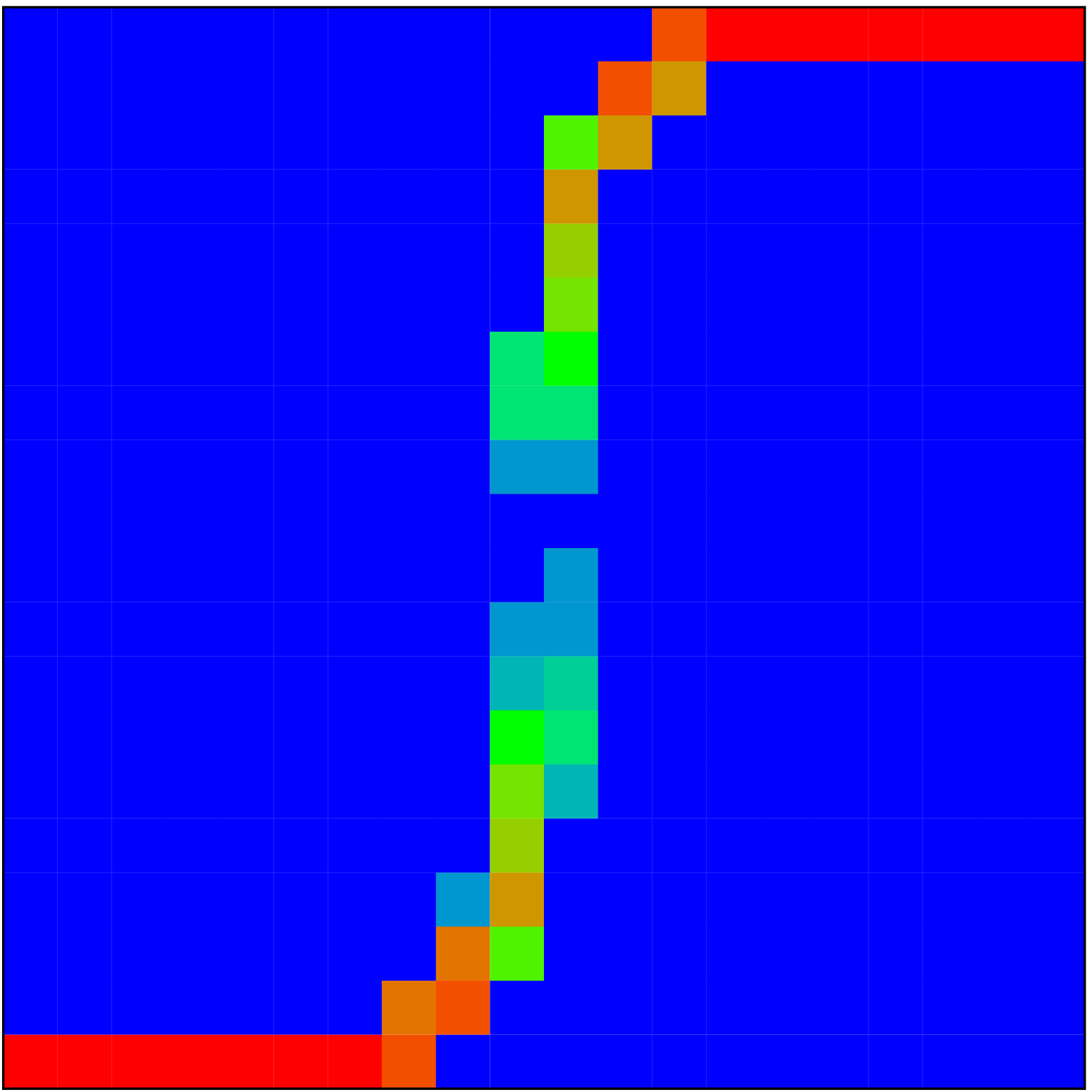}
\includegraphics[width=0.23\textwidth]{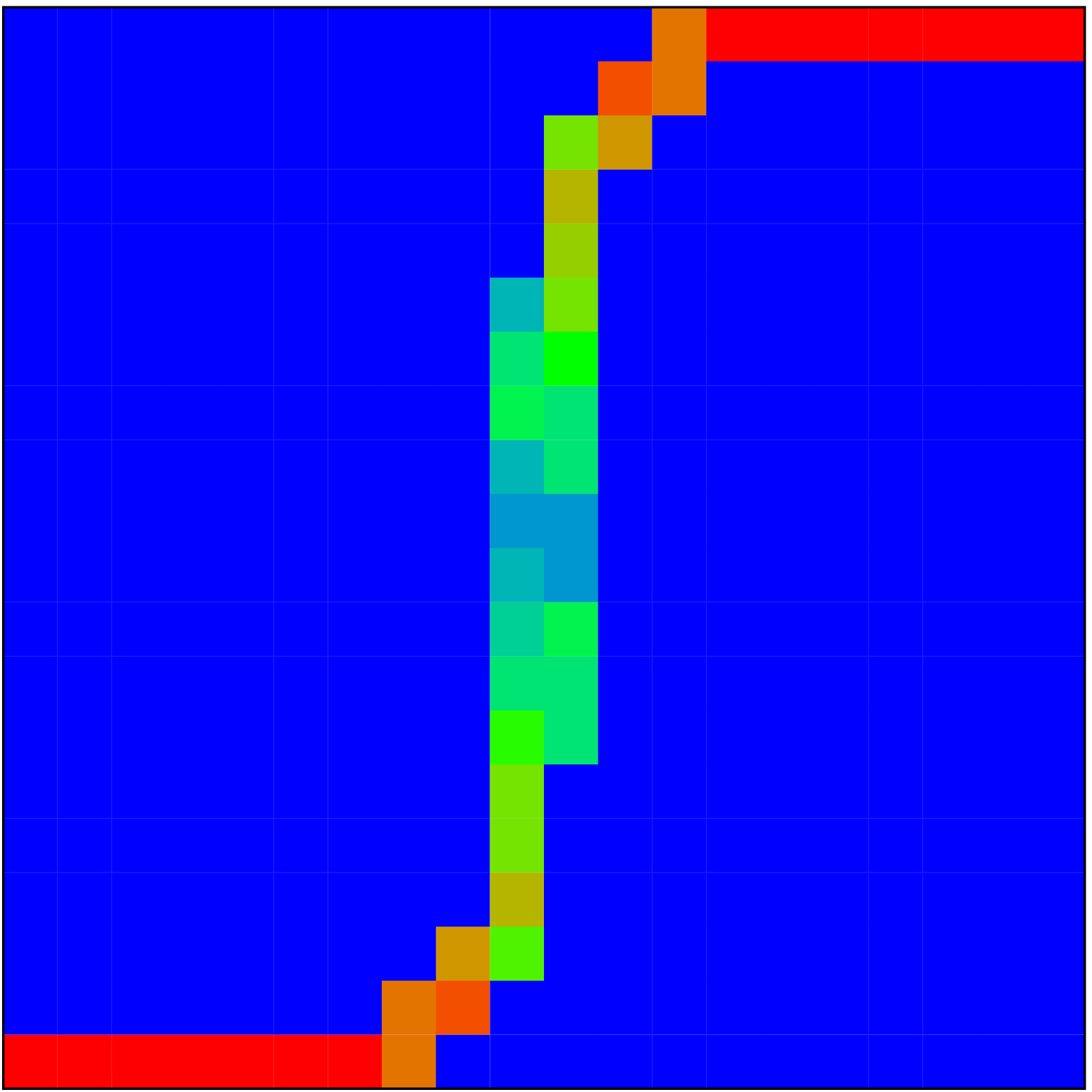}
\end{center}
\vglue -0.2cm
\caption{(Color online) Same as in Fig.~\ref{fig3}
(middle panels)
at $a=0.5$ but with uniform condition for spin flip being 
independent of PageRank probability (top panels: $P=1$ in Eq.~(\ref{eq1}))
and PageRank probability $P$ replaced by $\sqrt{P}$ in Eq.(\ref{eq1})
(bottom panels);
left and right panels correspond to Cambridge and Oxford networks;
here $T=0$ and $N_r=10^4$ realizations are used.
\label{fig5}} 
\end{figure}

Our interpretation of these results is the following.
For small values of $a \rightarrow 0$
the opinion of a given society member is
determined mainly by the PageRank of neighbors
{\it to whom he points to} (outgoing links).
The PageRank probability $P$ of nodes, on which many nodes point to,
is usually high since $P$ is proportional to the number of 
ingoing links \cite{meyerbook}.
Thus, at  $a \rightarrow 0$ a society  is composed of
members who form their opinion listening an elite opinion.
In such a society its elite
with one color opinion can impose
this opinion to a large fraction of the society.
This is illustrated on Fig.~\ref{fig4}
which shows a dependence of final fraction of red $f_f$ nodes
on parameter $a$ 
for a small initial fraction of red nodes
in the top values of PageRank index ($N_{top}=2000$).
We see that $a=0$ corresponds to a conformist society
which follows in its great majority the opinion of its elite.
For $a=1$ this fraction $f_f$ drops significantly showing that
this corresponds to a regime of tenacious society.
It is somewhat surprising that the   
tenacious society ($a \rightarrow 1$)
has well defined and relatively large fixed opinion phase with a
relatively small region of  bistability phase.
This is in a contrast to the conformist society
at $a \rightarrow 0$ when the opinion is strongly influenced 
by the society elite. We attribute this to the fact 
that in Fig.~\ref{fig3} we start with a randomly 
distributed opinion, due to that the opinion of elite
has two fractions of two colors that 
create a bistable situation since two fractions
of society follows opinion of this divided elite
that makes the situation bistable on a larger
interval of $f_i$ compared to the case of tenacious society
at $a \rightarrow 1$.
 
To stress the important role of PageRank in the 
dependence of $f_f$ on $f_i$ presented in Fig.~\ref{fig3}
we show in Fig.~\ref{fig5} the same analysis at $a=0.5$
but for the case when in Eq.(\ref{eq1}) for the spin flip
we take all $P=1$ (equal weight for all nodes). 
The data of Fig.~\ref{fig5}
clearly demonstrate that in this case the bistability
of opinion disappears. Thus the PROF model is qualitatively
different from the case when only the links without their
PageRank weight are counted for the spin flip condition.
We also test the sensitivity in respect to PageRank probability
replacing $P$ by $\sqrt{P}$ in Eq.(\ref{eq1})
as it is shown in Fig.~\ref{fig5} (bottom panels).
We see that compared to the case $P=1$ we start to have
some signs of bistability but still they remain rather weak
compared to the case of Fig.~\ref{fig3}.  

\begin{figure}[ht]
\begin{center}
\includegraphics[width=0.23\textwidth]{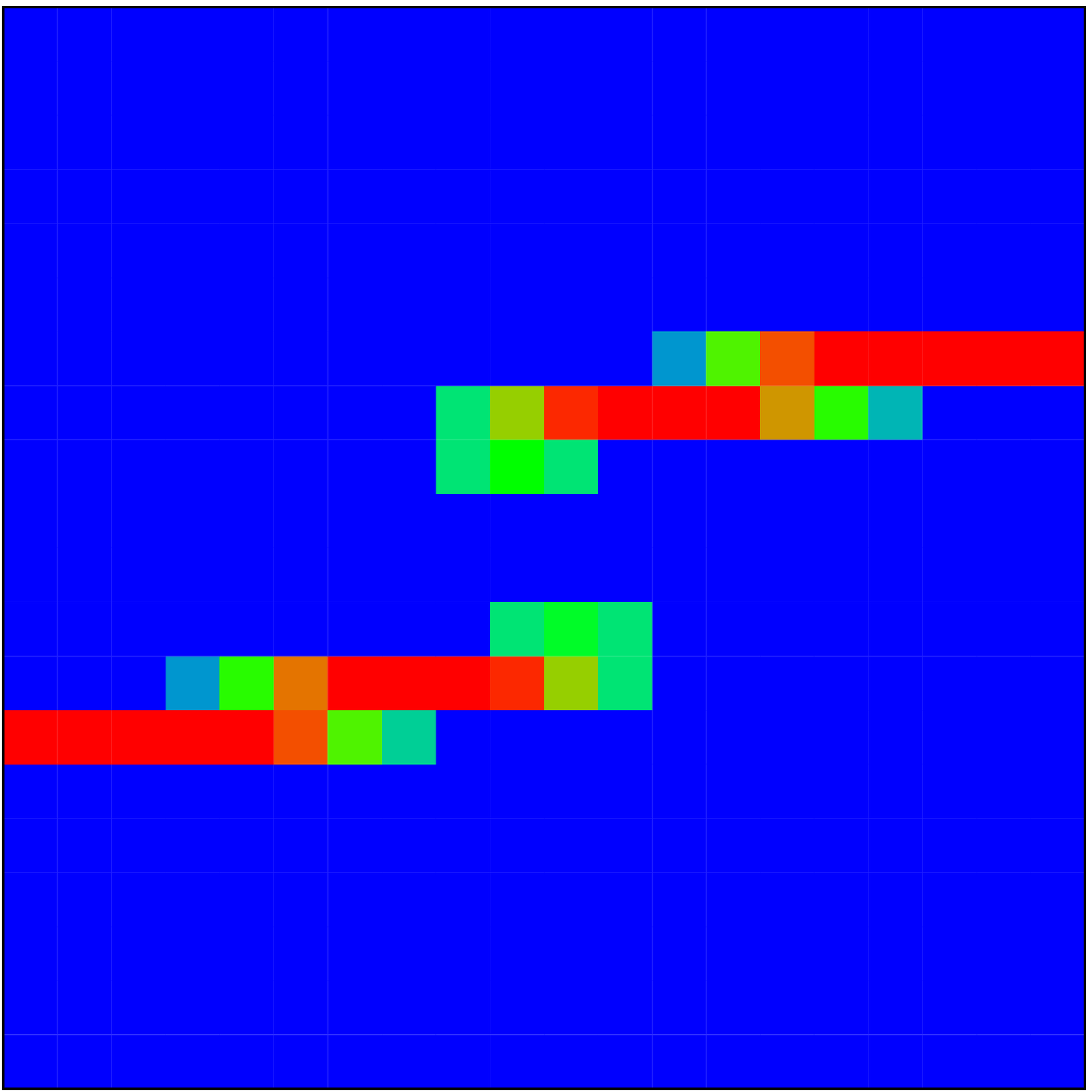}
\includegraphics[width=0.23\textwidth]{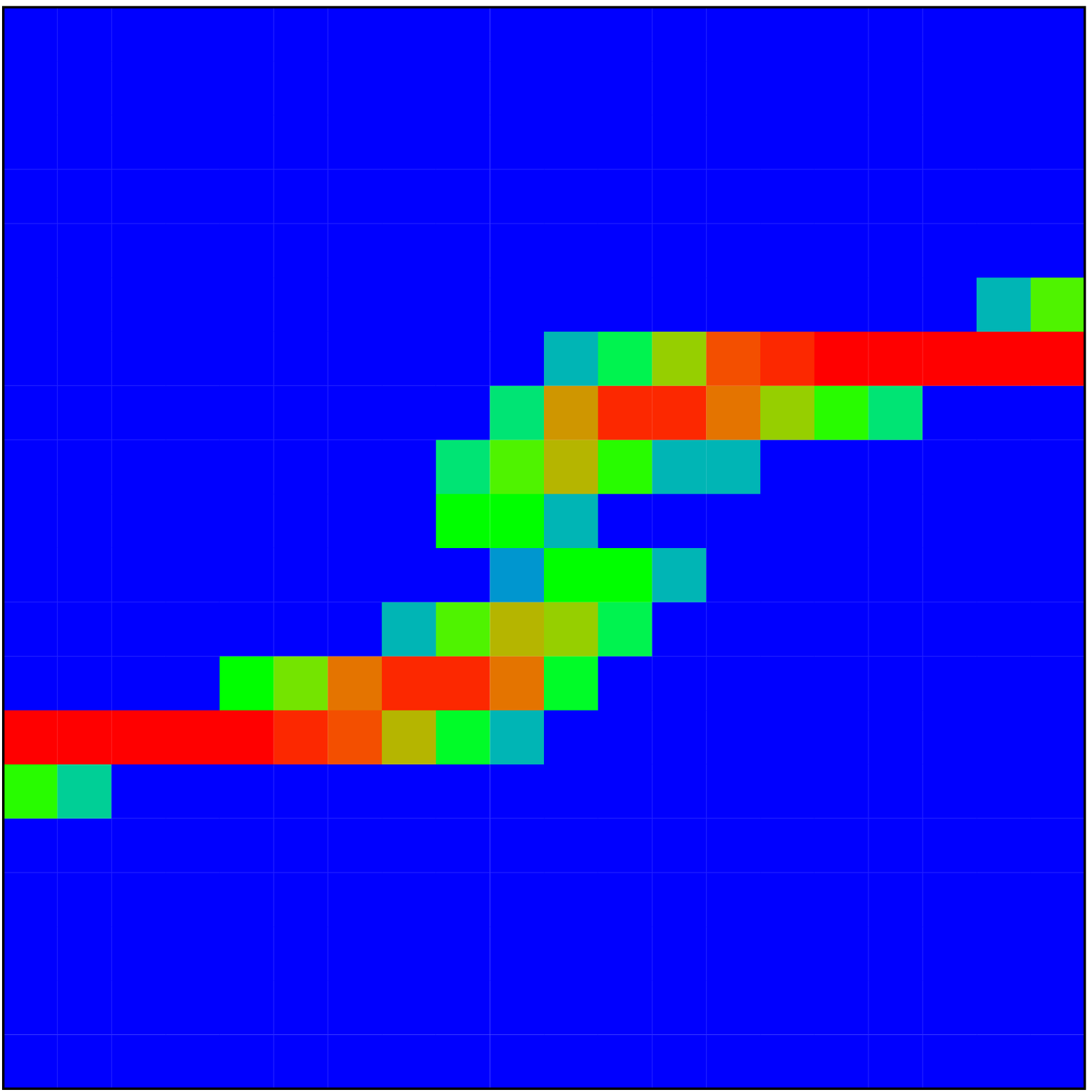}\\
\includegraphics[width=0.23\textwidth]{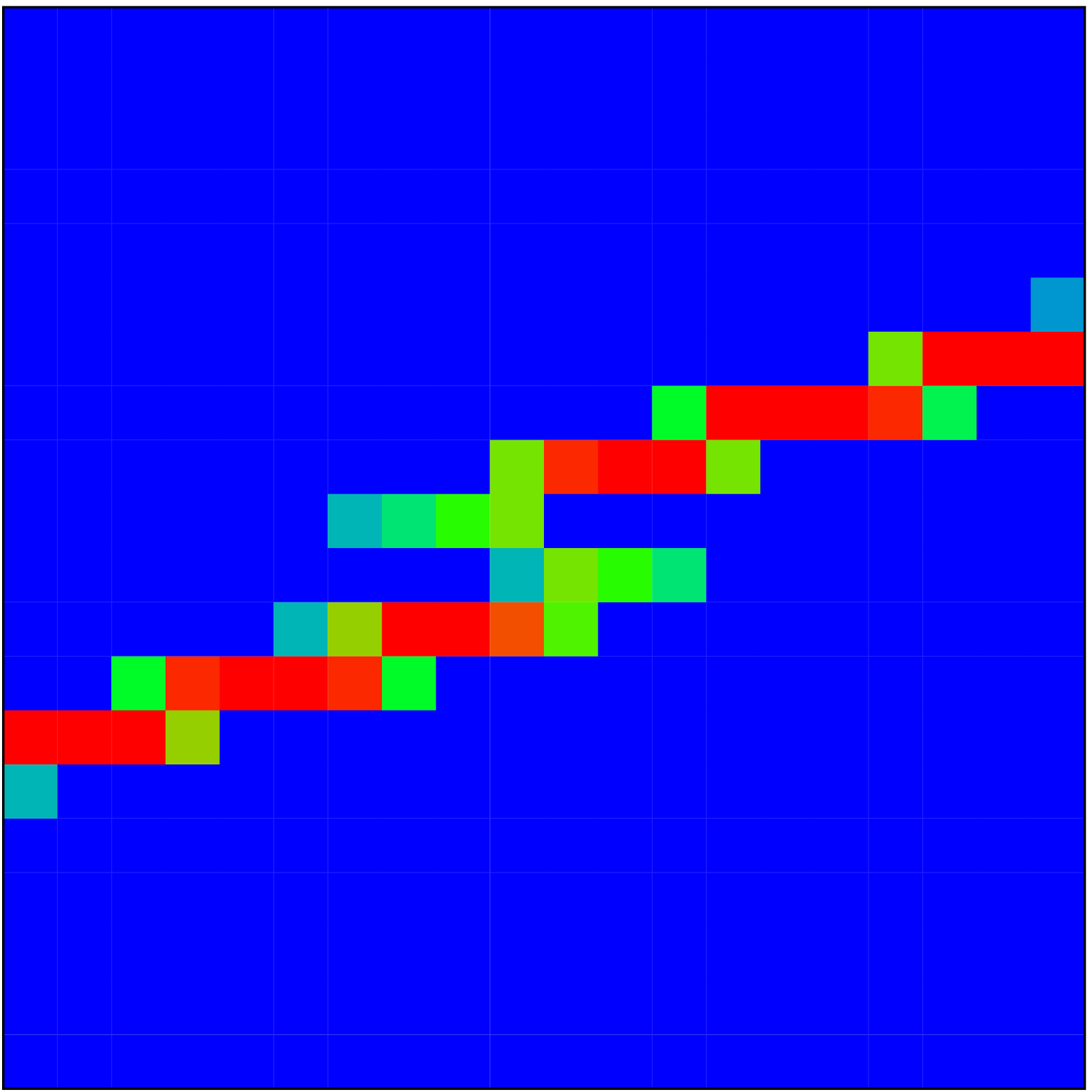}
\includegraphics[width=0.23\textwidth]{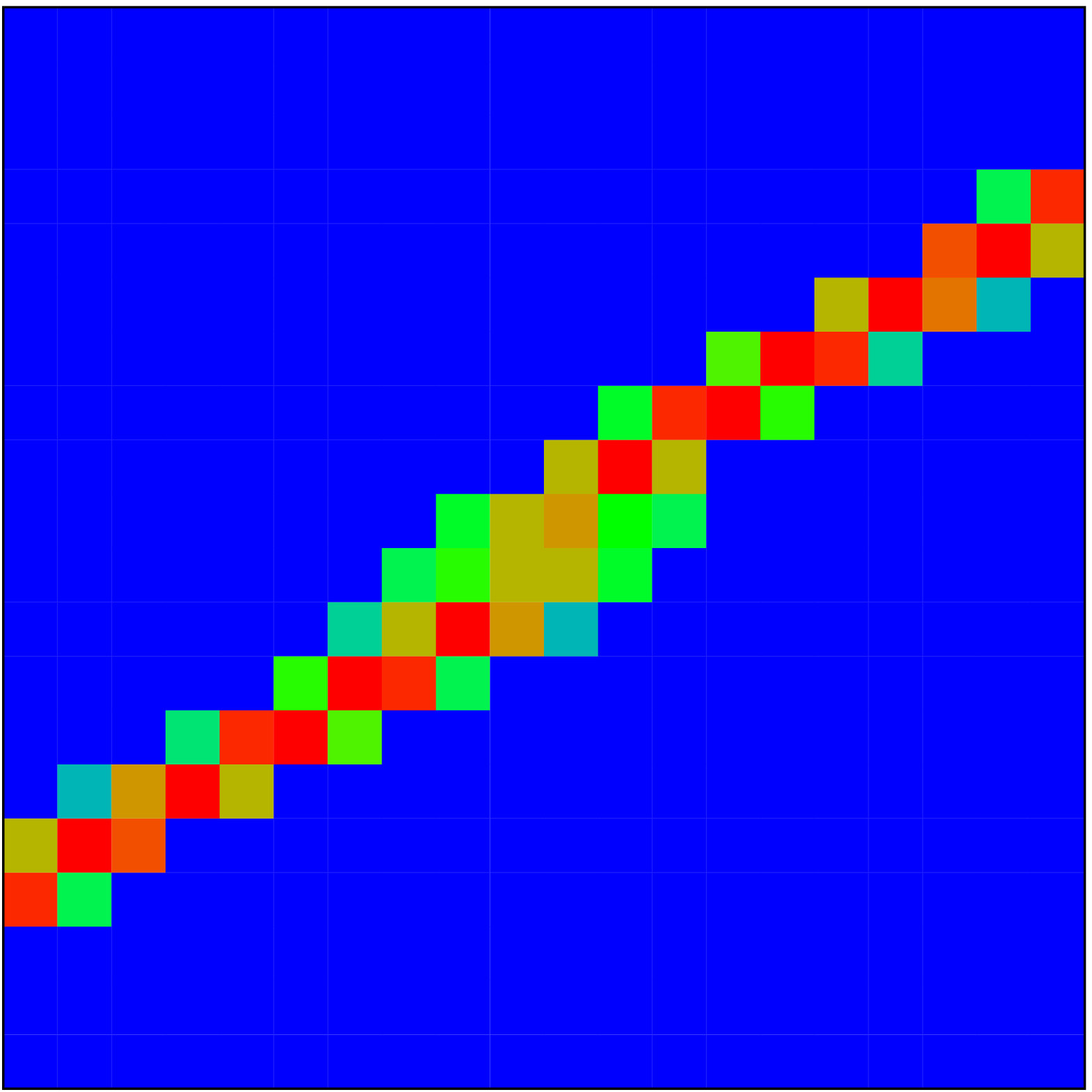}
\end{center}
\vglue -0.2cm
\caption{(Color online) Same as in Fig.~\ref{fig3} (middle panel)
at $a=0.5$ but at finite temperature
$T$ during the relaxation process with
$T=0.001$ (top panels) and $T=0.01$ (bottom panels);
the number of random initial realizations is
$N_r=6000$, the relaxation is done
during $t=200$ iterations.
Left and right columns correspond to
Cambridge and Oxford networks.
\label{fig6}} 
\end{figure}

In fact the spin flip condition (\ref{eq1})
can be viewed as a relaxation process in a disordered ferromagnet
(since all $J_{ij} \geq 0$ in (\ref{eq2}), (\ref{eq3}))
at zero temperature. Such type of analysis of voter model relaxation 
process on regular lattices is analyzed 
in \cite{fortunatormp,krapivskybook}.
From this view point it is natural to consider the effect of 
finite temperature $T$ on this relaxation.
At finite $T$ the flip condition is determined by the 
thermal Metropolis probability $\exp(-\Delta \epsilon_i/T)$
as described in previous Section. We follow this thermodynamic 
relaxation process at finite temperature up to $t=200$ iterations
and in this way obtain the probability distribution 
of final $f_f$ fraction of red nodes obtained from
initial $f_i$ fraction of red nodes
randomly distributed over the network at $t=0$.
The results obtained at finite temperatures are shown at Fig.~\ref{fig6}.
They show that a finite temperature $T$   allows to have 
a finite fraction of red nodes $f_f$ when for their 
small initial fraction $f_i$ all final
$f_f$ were equal to zero. Also the bistability splitting is reduced
and it disappears at larger values of $T$.
Thus finite $T$ introduce a certain smoothing in 
$W_f$ distribution.
\begin{figure}[ht]
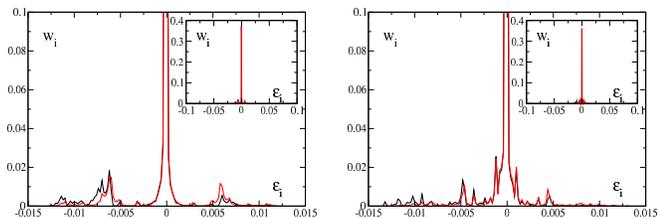

\begin{center}
\includegraphics[width=0.23\textwidth]{fig7a}
{\hskip 0.2cm}
\includegraphics[width=0.23\textwidth]{fig7b}
\end{center}
\vglue -0.2cm
\caption{(color online) Normalized histograms of 
probability distribution $w_i$
over local energies $\epsilon_i$ obtained from the relaxation process
during $t=10^3$ time iterations  
at temperatures $T=0.01$ (black curve) and $T=0.05$ (red/gray curve);
average is done over  $N_r =200$ random initial realizations.  
The insets show the distributions on a large scale
including all local energies $\epsilon_i$. 
Left and right panels show Cambridge and Oxford networks.
\label{fig7}}
\end{figure}

However, the relaxation process at finite temperatures does not lead to
the thermal Boltzmann distribution. Indeed, in Fig.~\ref{fig7}
we show the probability distribution $w_i(\epsilon_i)$
as a function of local energies $\epsilon_i$ 
defined in (\ref{eq2}), (\ref{eq3}).
The distribution $w_i(\epsilon_i)$ is
obtained from the relaxation process with many initial 
random spin realizations $N_r$.  Even if the temperature $T$
is comparable with typical values of local energies $\epsilon_i$
we still obtain a rather peaked distribution at
$\epsilon_i \approx 0$ being very different from the Boltzmann distribution.
\begin{figure}[ht]
\begin{center}
\includegraphics[width=0.23\textwidth]{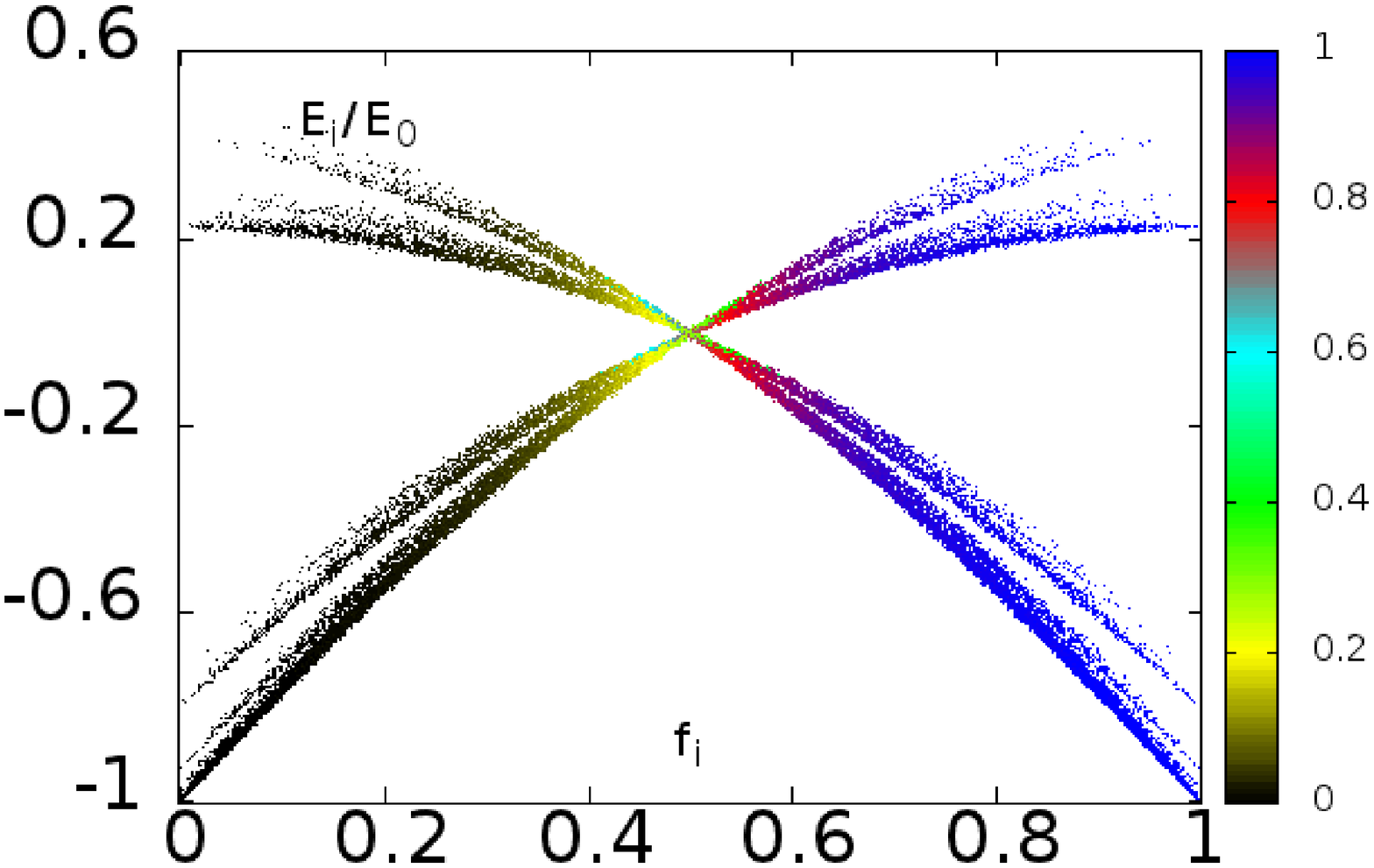}
\includegraphics[width=0.23\textwidth]{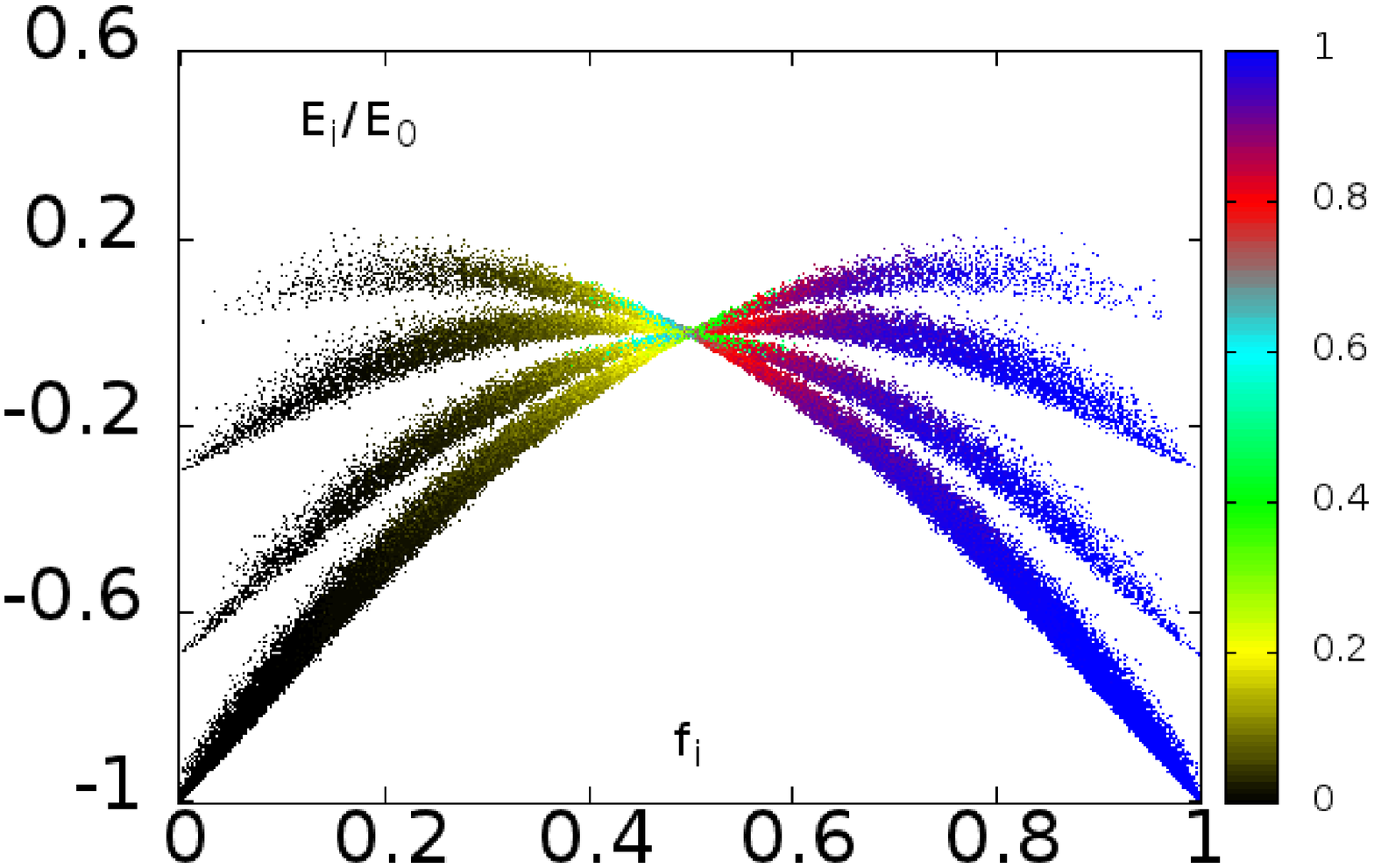}
\end{center}
\vglue -0.2cm
\caption{(Color online) Diagram  shows the final fraction of red nodes $f_f$,
coded by color from $f_f=0$ (black) to $f_f=1$ (blue/dark gray),
as a function of initial fraction of red nodes $f_i$
and the total initial energy $E_i$; each of $N_r > 3.5 \times 10^4$ random
realizations is shown by color point; 
data are shown after $t=20$ time iterations
at $T=0$.
The energy $E_0$ is the modulus of 
total energy with all spin
up; here $a=0.5$. Left and right panels show data for
Cambridge ($E_0=341.20$) and Oxford ($E_0=254.28$) networks;
bars show color attribution to final probability $f_f$.
\label{fig8}}
\end{figure}

We argue that a physical reason of significantly non-Boltzmann
distribution is related to the local nature of spin flip condition which
does not allow to produce a good thermalization on a scale of the whole system.
Indeed, there are various energetic branches and probably nonlocal
thermalization flips of group of spins are required for a better thermalization.
However, the voting is a local process that involves only direct neighbors
that seems to be not sufficient for emergence of 
the global thermal distribution.
The presence of a few energy branches is well visible from the data
of Fig.~\ref{fig8} obtained at $T=0$. This figure 
shows the diagram of final fraction $f_f$ of red nodes
in dependence on their initial fraction $f_i$ and the total 
initial energy $E_i= \sum_{m=1}^N \epsilon_m$ of the whole system
corresponding to a chosen initial random configuration of spins.
Most probably, these different branches
prevent efficient thermalization of the system
with only local spin flip procedure.
In addition to the above points the asymmetric form of $J_{ij}$
couplings plays an important role generating
more complicated picture compared to the usual
image of thermal relaxation (see e.g. \cite{galam1}).
We also note that the thermalization is 
absent in voter models
on regular lattices \cite{fortunatormp}.

\section{IV. PROF-Sznajd model}
The Sznajd model \cite{sznajd} nicely incorporates the 
well-known trade union principle ``United we stand, divided we fall''
into the field of voter modeling and opinion formation on regular networks.
The review of various aspects of this model is given in \cite{fortunatormp}.
Here we generalize the Sznajd model to include in it the
features of PROF model and consider it on social networks
with their scale-free structure. This gives us the PROF-Sznajd model 
which is constructed in the following way. For a given network we
determine the PageRank probability $P(K_i)$ and PageRank index $K_i$
for all $i$ nodes. After that we introduce the definition of 
{\it group} of nodes. It is defined by the following rule applied 
at each time step $\tau$:
\begin{itemize}
\item {\it i)} pick by random a node $i$ in the network and 
consider the polarization of the $N_g-1$ highest PageRank nodes pointing to it;
\item {\it ii)} if the node $i$ and all other $N_g-1$ nodes have the same color
(same spin polarization), then these $N_g$ nodes form 
a group whose effective PageRank value is 
the sum of all the member values $P_{g}=\sum_{j=1}^{N_g} P_j$; 
if it is not the case then we leave the nodes unchanged 
and perform the next time step;
\item {\it iii)} consider all the nodes 
{\it pointing to any member of the group} 
(this corresponds to the model \emph{option 1}) 
or consider {\it all the nodes 
pointing to any member of the group 
and all the nodes pointed by any member of the group} 
(this corresponds to the model \emph{option 2});
then check all these nodes $n$ directly linked to the group:
if an individual node PageRank value $P_n$  is less than $P_{group}$ then
this node joins the group by taking the same color (polarization) 
as the group nodes; 
if it is not the case then a node is left unchanged; 
the PageRank values of 
added nodes are then added to the group PageRank $P_{group}$ 
and the group size is increased.
\end{itemize}
The above time step is repeated many times during time $\tau$,
counting the number of steps,
by choosing a random node $i$ on each next  step. 
This procedure effectively
corresponds to the zero temperature case in the PROF model.
\begin{figure}[ht]
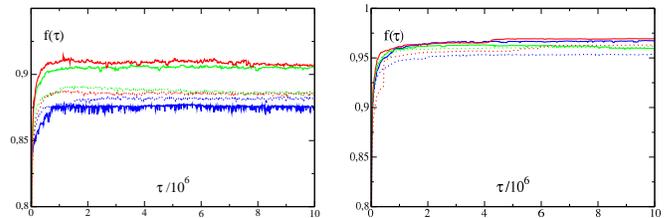

\vglue +0.2cm
\begin{center}
\includegraphics[width=0.23\textwidth]{fig9a}
{\hskip 0.2cm}
\includegraphics[width=0.23\textwidth]{fig9b}
\end{center}
\vglue -0.2cm
\caption{(Color online) Time evolution of the fraction of red nodes
$f(\tau)$ in the PROF-Sznajd model with the initial fraction of
red nodes $f_i=0.7$ at one random realization. 
The curves show data for three values of 
group size $N_g=3$ (blue/black); $8$ (green/light gray);
$13$ (red/gray). Full/dashed curves are for Cambridge/Oxford networks;
left panel is for option 1; right panel is for option 2. 
\label{fig9}}
\end{figure}

\begin{figure}[ht]
\begin{center}
\includegraphics[width=0.23\textwidth]{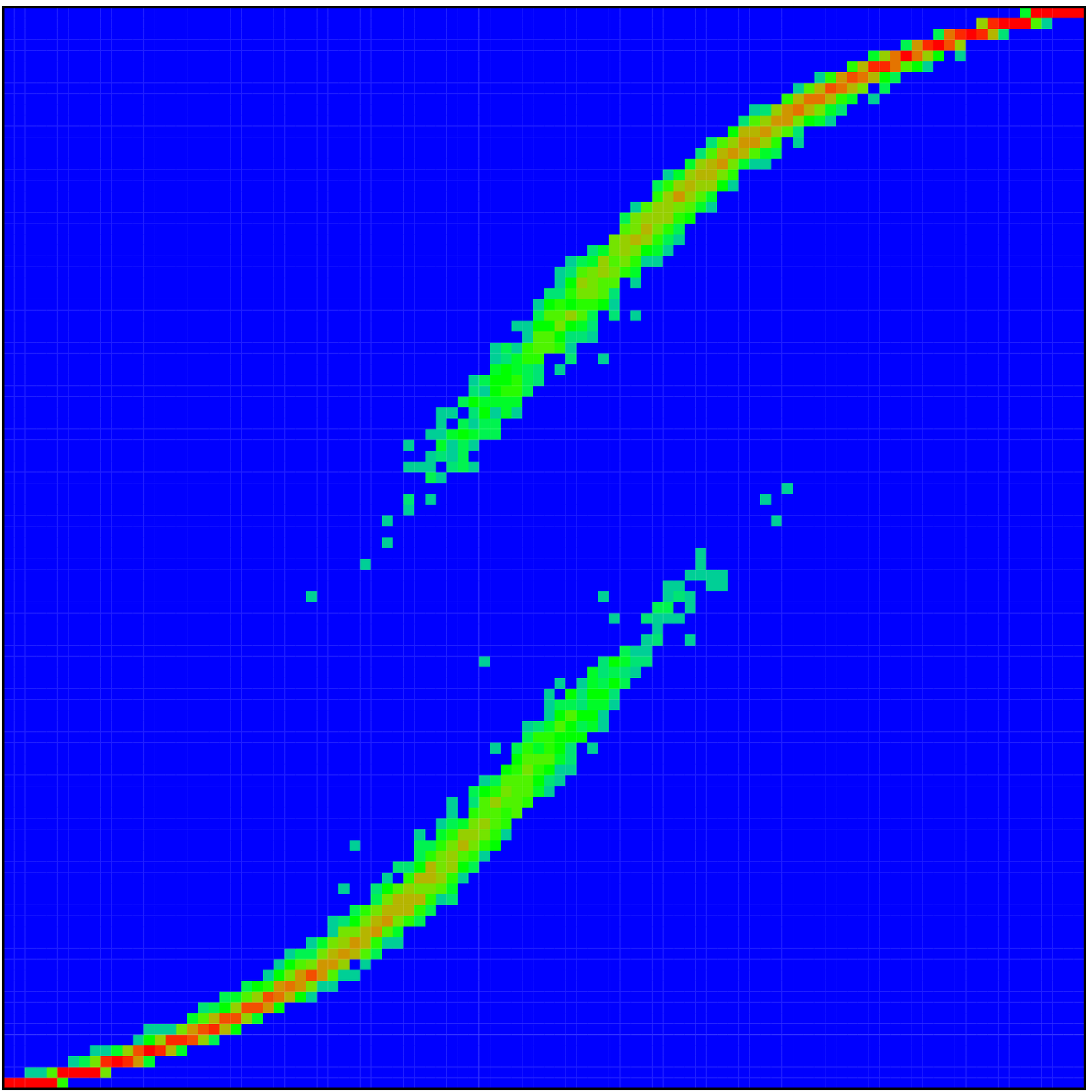}
\includegraphics[width=0.23\textwidth]{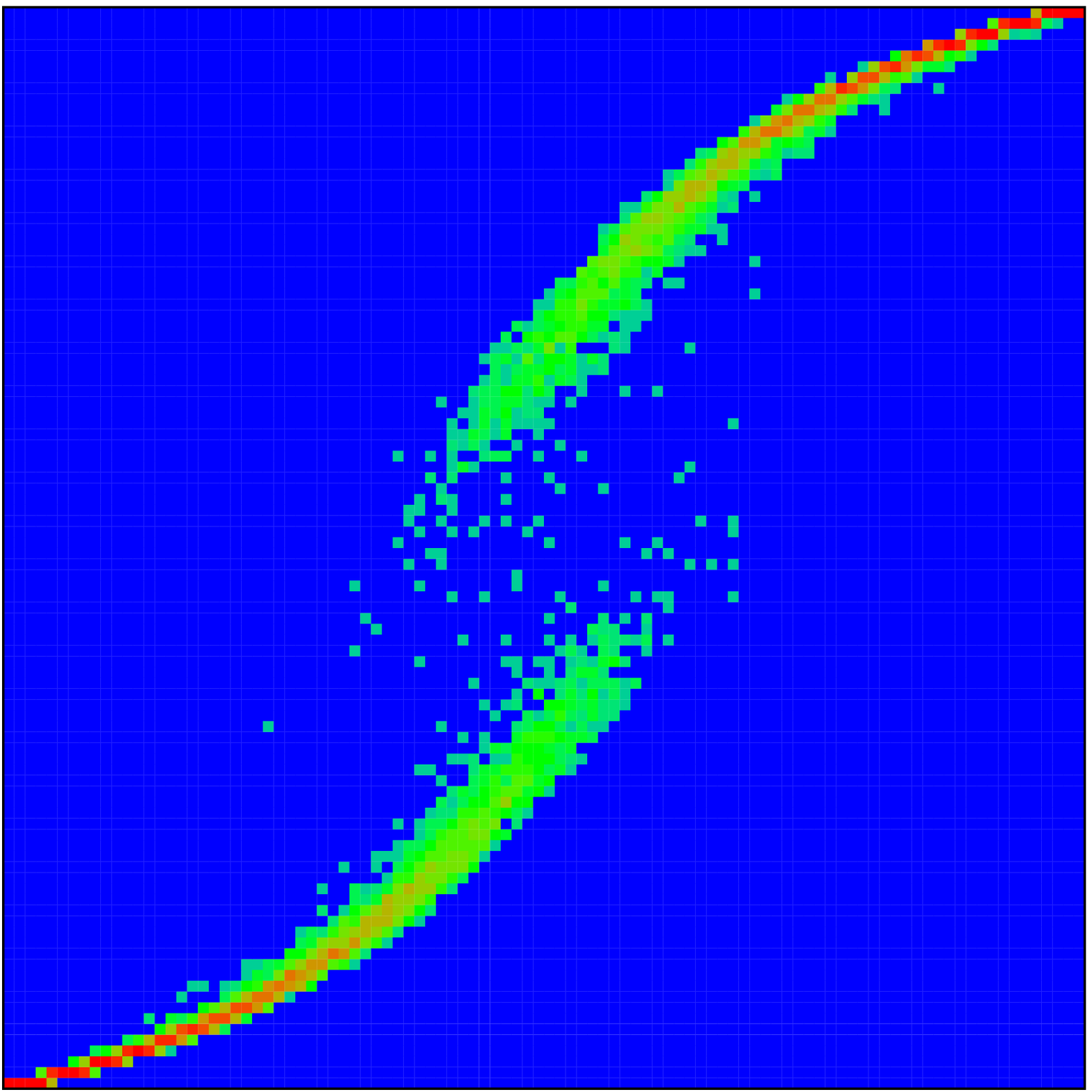}\\
\includegraphics[width=0.23\textwidth]{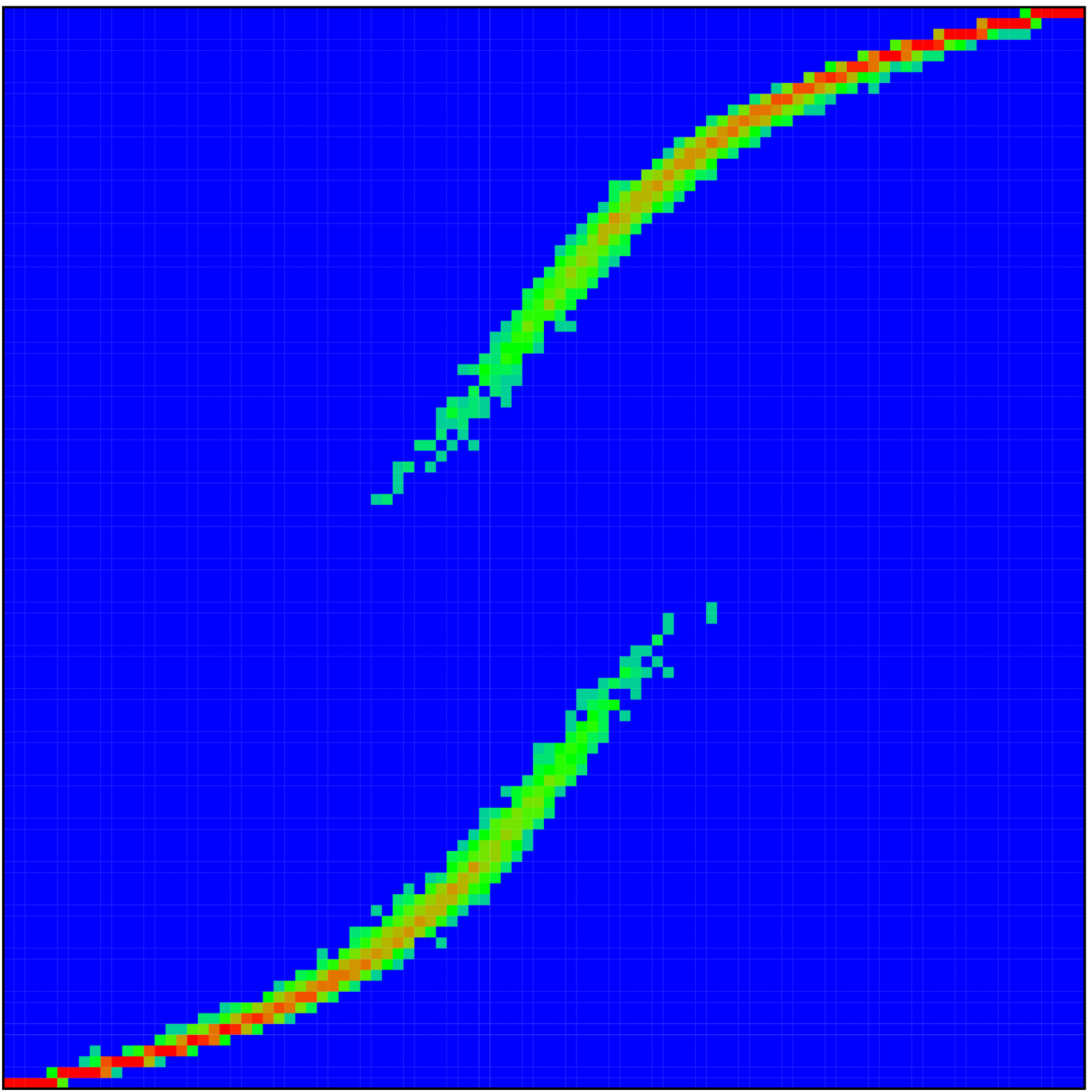}
\includegraphics[width=0.23\textwidth]{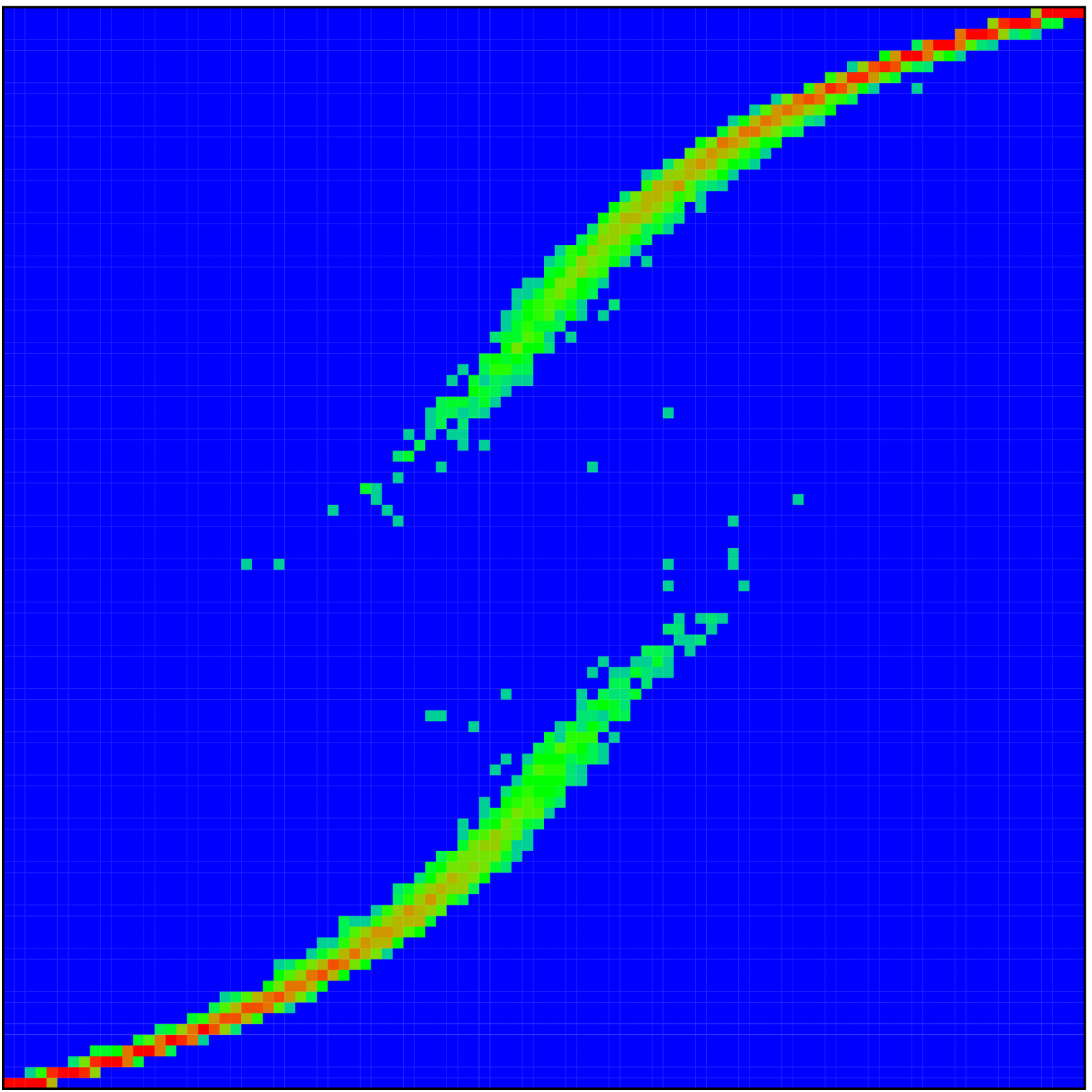}\\
\includegraphics[width=0.23\textwidth]{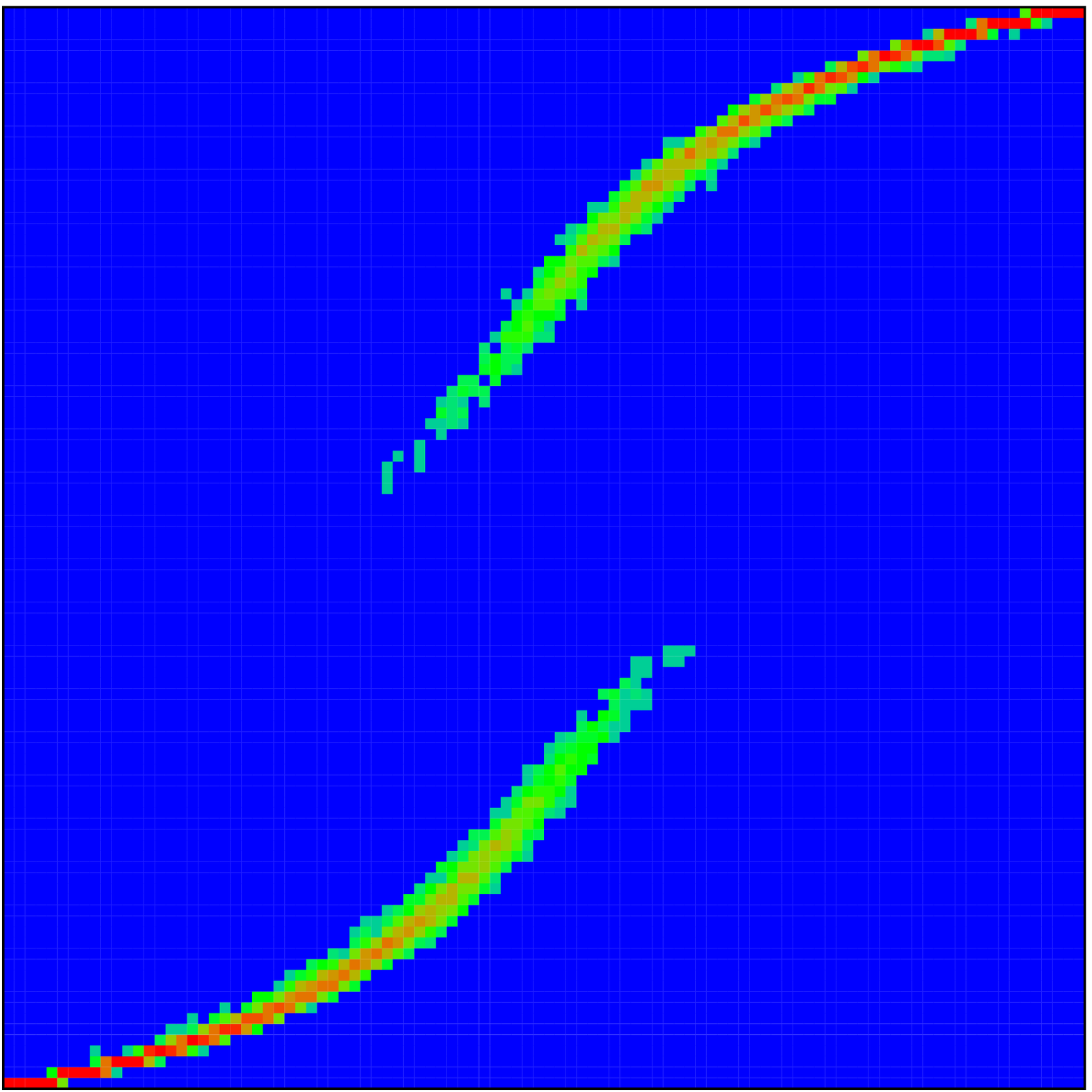}
\includegraphics[width=0.23\textwidth]{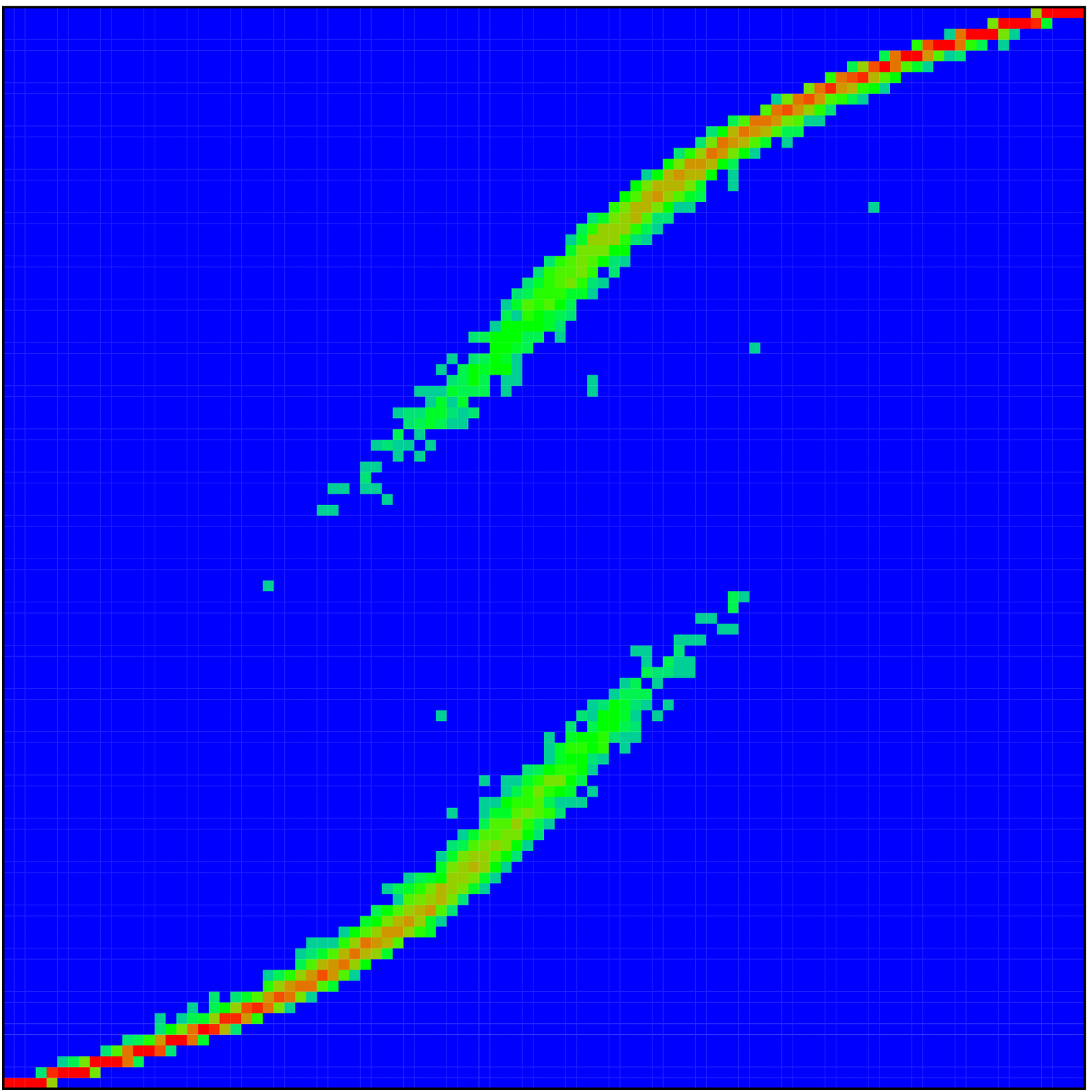}
\end{center}
\vglue -0.2cm
\caption{(Color online) PROF-Sznajd model, option 1:
density plot of probability $W_f$ 
to find a final red fraction $f_f$, shown in $y-$axis,
in dependence on an initial red fraction $f_i$, shown in $x-$ axis;
data are shown inside the unit square $0\leq f_i,f_f \leq 1$.
The values of $W_f$ are defined as a relative number of
realizations found inside each of $100\times100$ cells
which cover  the whole unit square. Here $N_r=10^4$ realizations of
randomly distributed colors are used to obtained $W_f$ values;
for each realization the
time evolution is followed up the convergence time with up to 
$\tau=10^7$ steps.
{\it Left column:} Cambridge network;
{\it right column:} Oxford network;
here $N_g=3, 8, 13$ from top to bottom.
The probability $W_f$ is proportional to
color changing from zero (blue/black) to unity
(red/gray).
\label{fig10}} 
\end{figure}

A typical example of the time evolution 
of the fraction of red nodes $f(\tau)$ in the PROF-Sznajd model 
is shown in Fig.~\ref{fig9}. It shows that the system converges
to a steady-state after a time scale $\tau_c \approx 10 N$
that is comparable with the convergence times for the PROF models
studied in previous Sections. We see that there are 
still some fluctuations in the steady-state regime 
which are visibly smaller for the option 2 case.
We attribute this to a larger number of direct links 
in this case. The  number of group nodes $N_g$
gives some variation of $f_f$ but these variations
remain on a relatively small scale of a few percents.
Here, we should point on the important difference 
between PROF and PROF-Sznajd models: 
for a given initial
color realization,
in the first case we have convergence to a fixed state
after some convergence time while in the second case
we have convergence to a steady-state which
continue to fluctuate in time, keeping 
the colors distribution only on average.

\begin{figure}[ht]
\begin{center}
\includegraphics[width=0.23\textwidth]{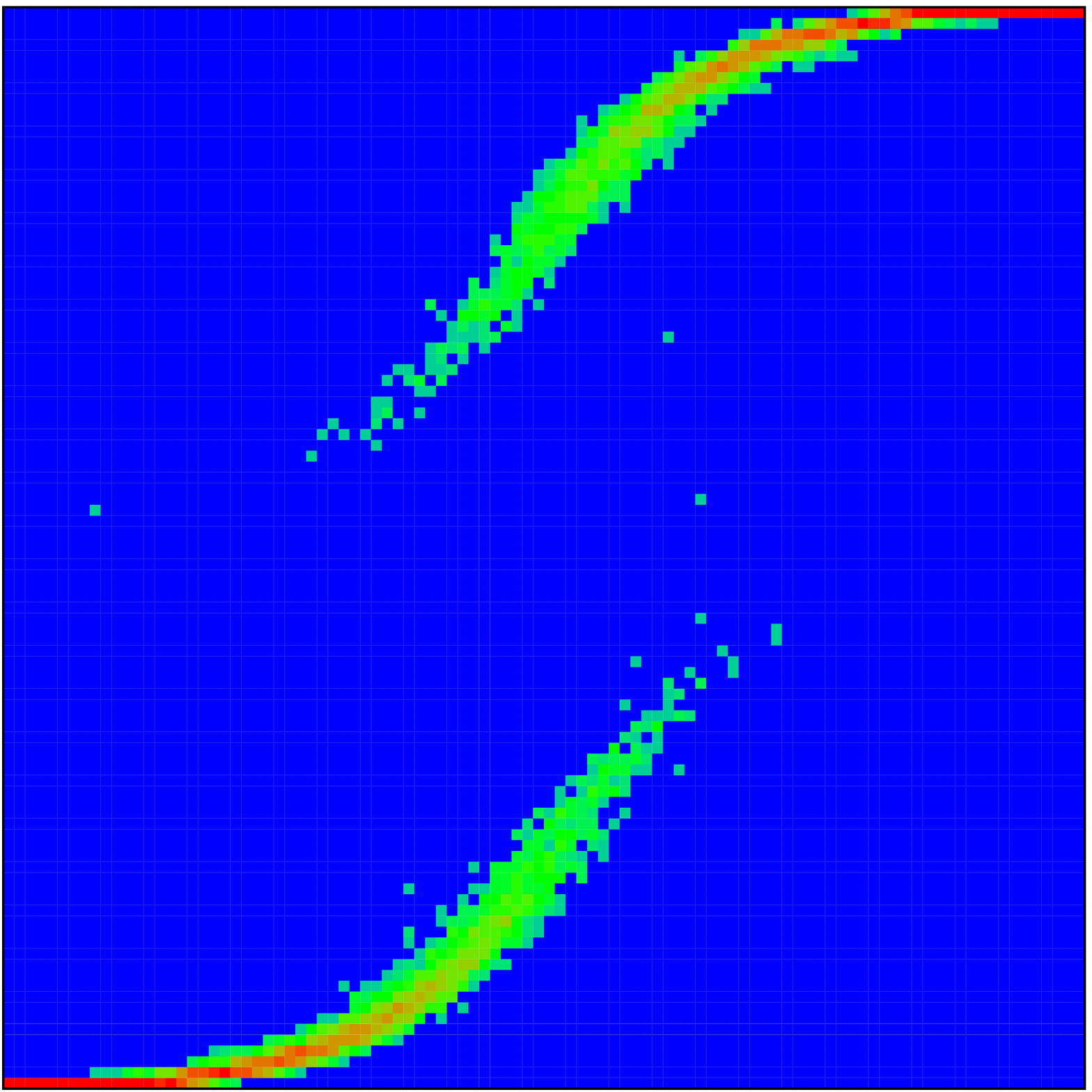}
\includegraphics[width=0.23\textwidth]{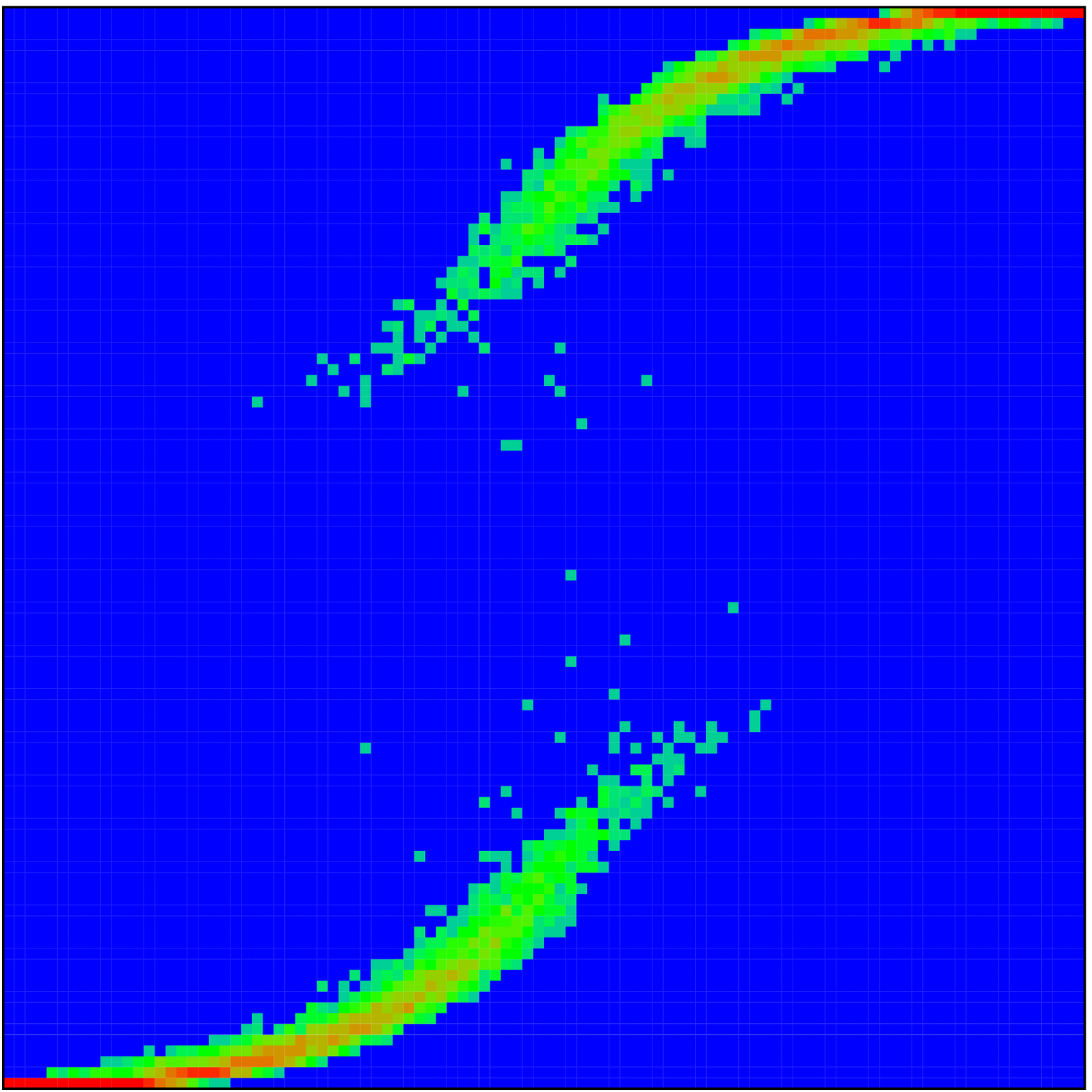}\\
\includegraphics[width=0.23\textwidth]{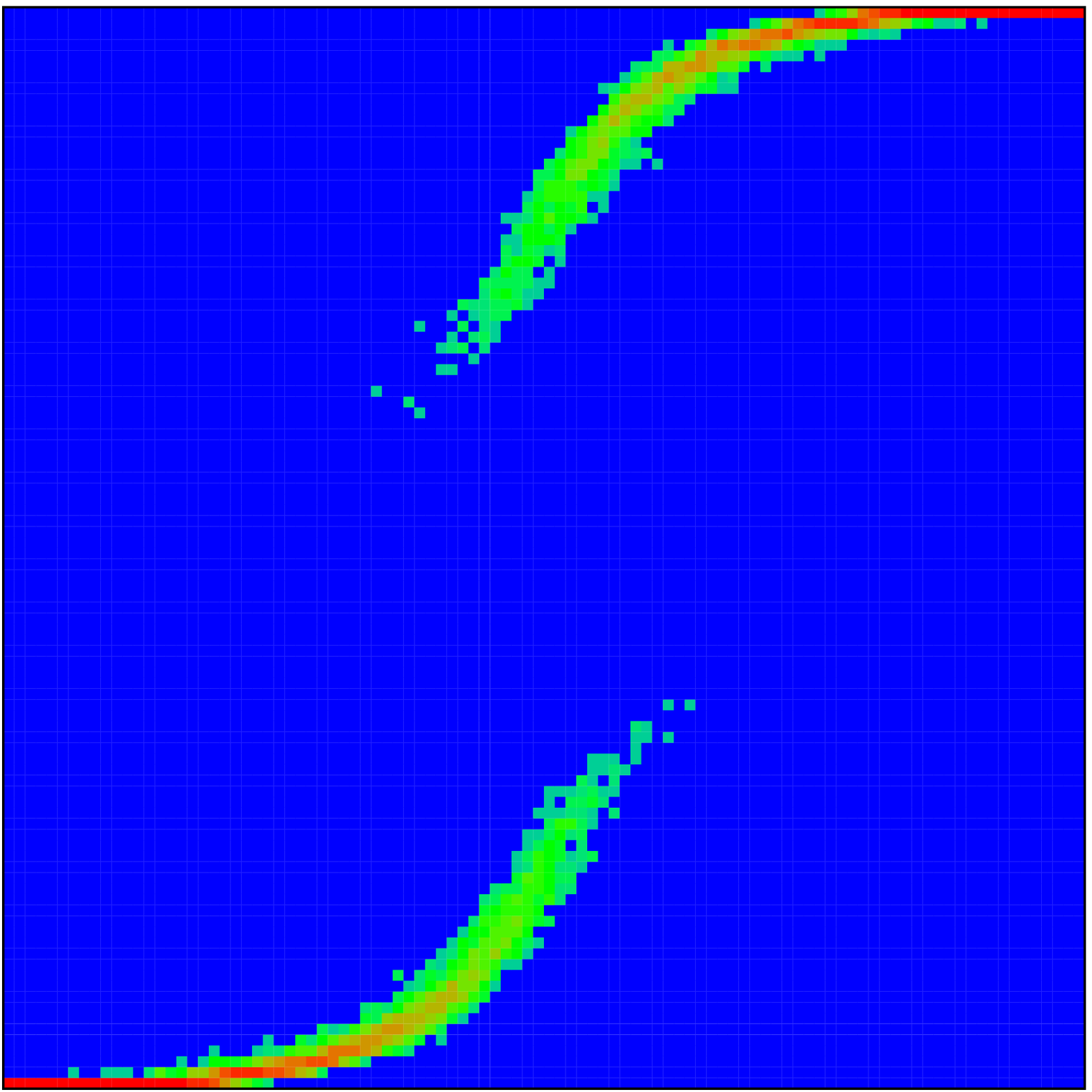}
\includegraphics[width=0.23\textwidth]{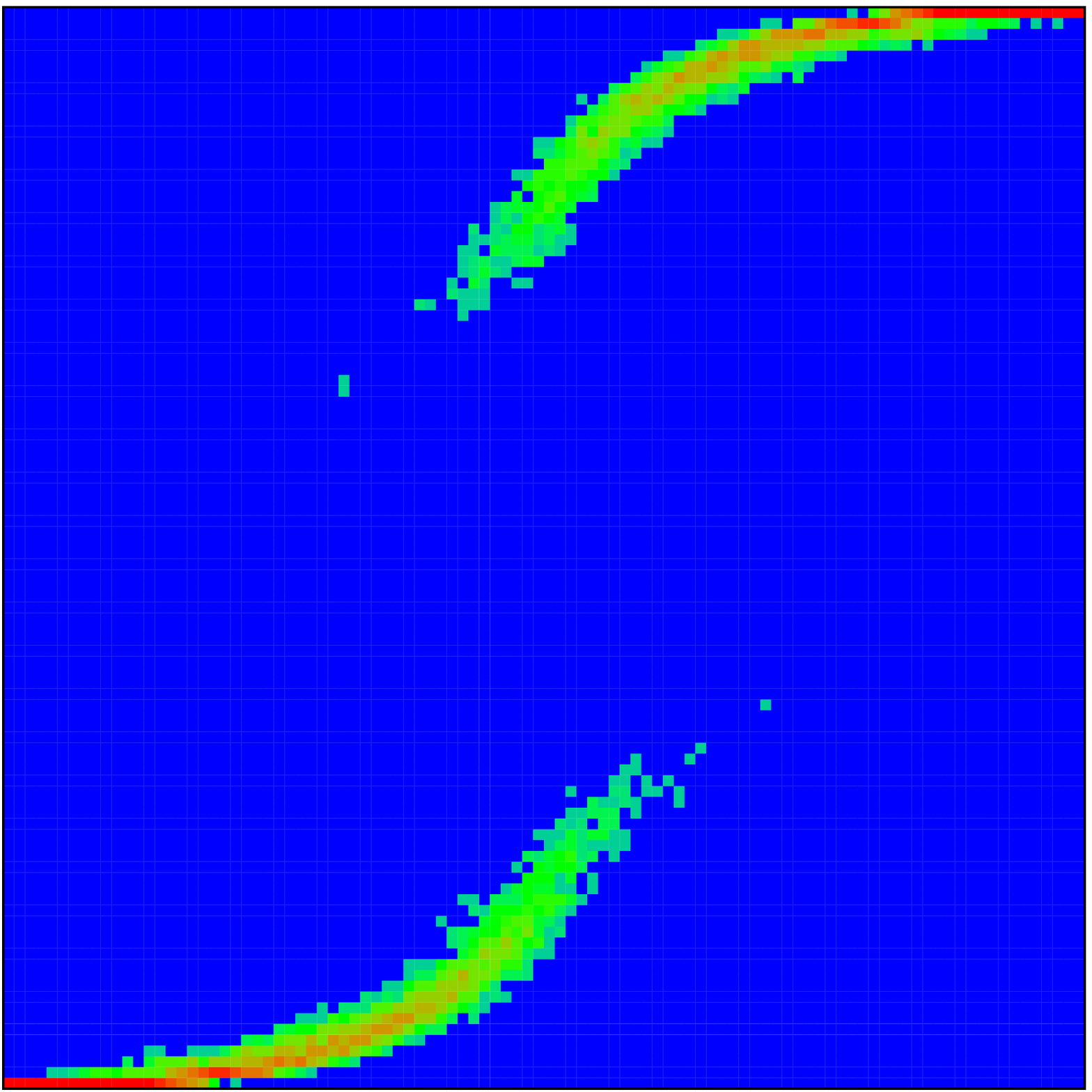}\\
\includegraphics[width=0.23\textwidth]{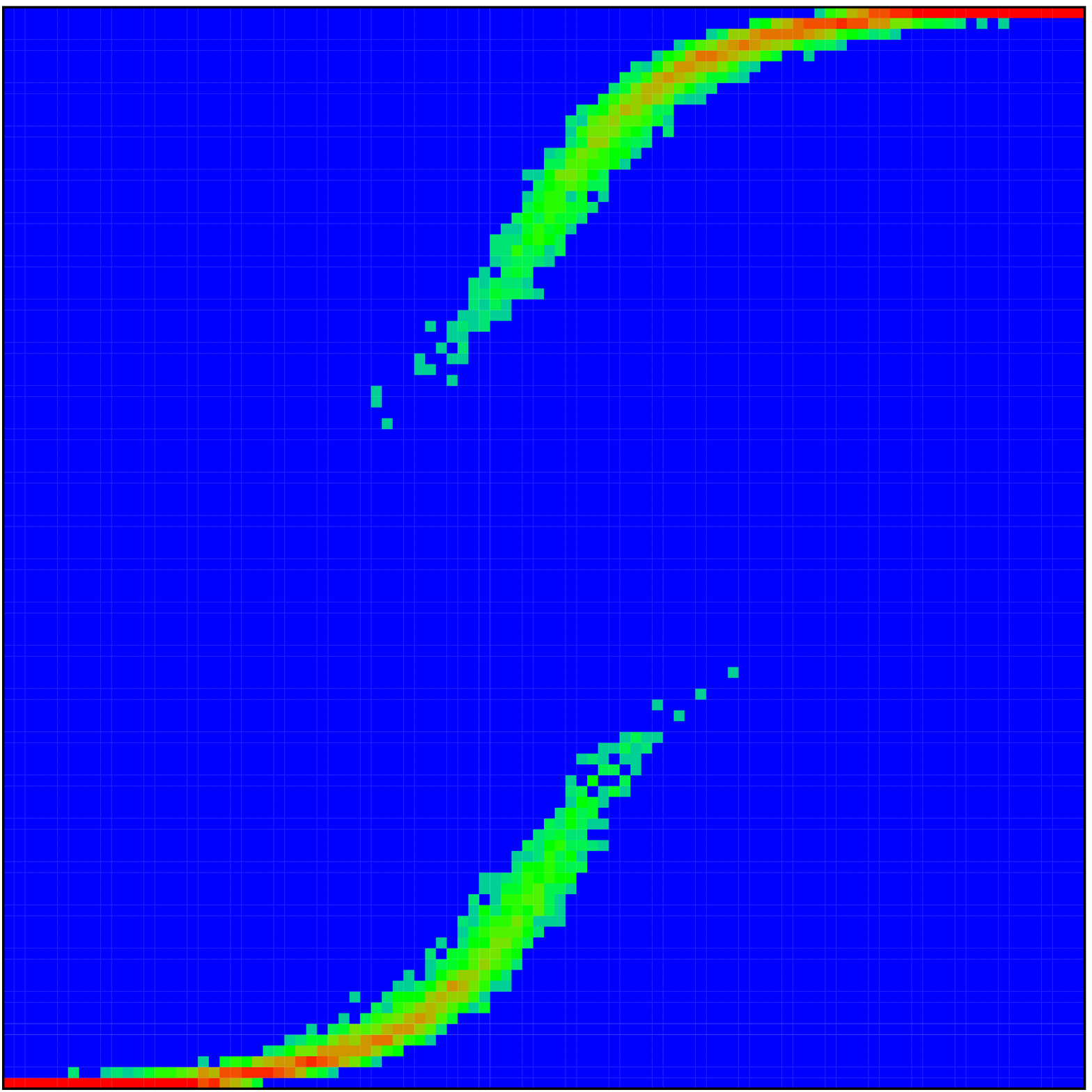}
\includegraphics[width=0.23\textwidth]{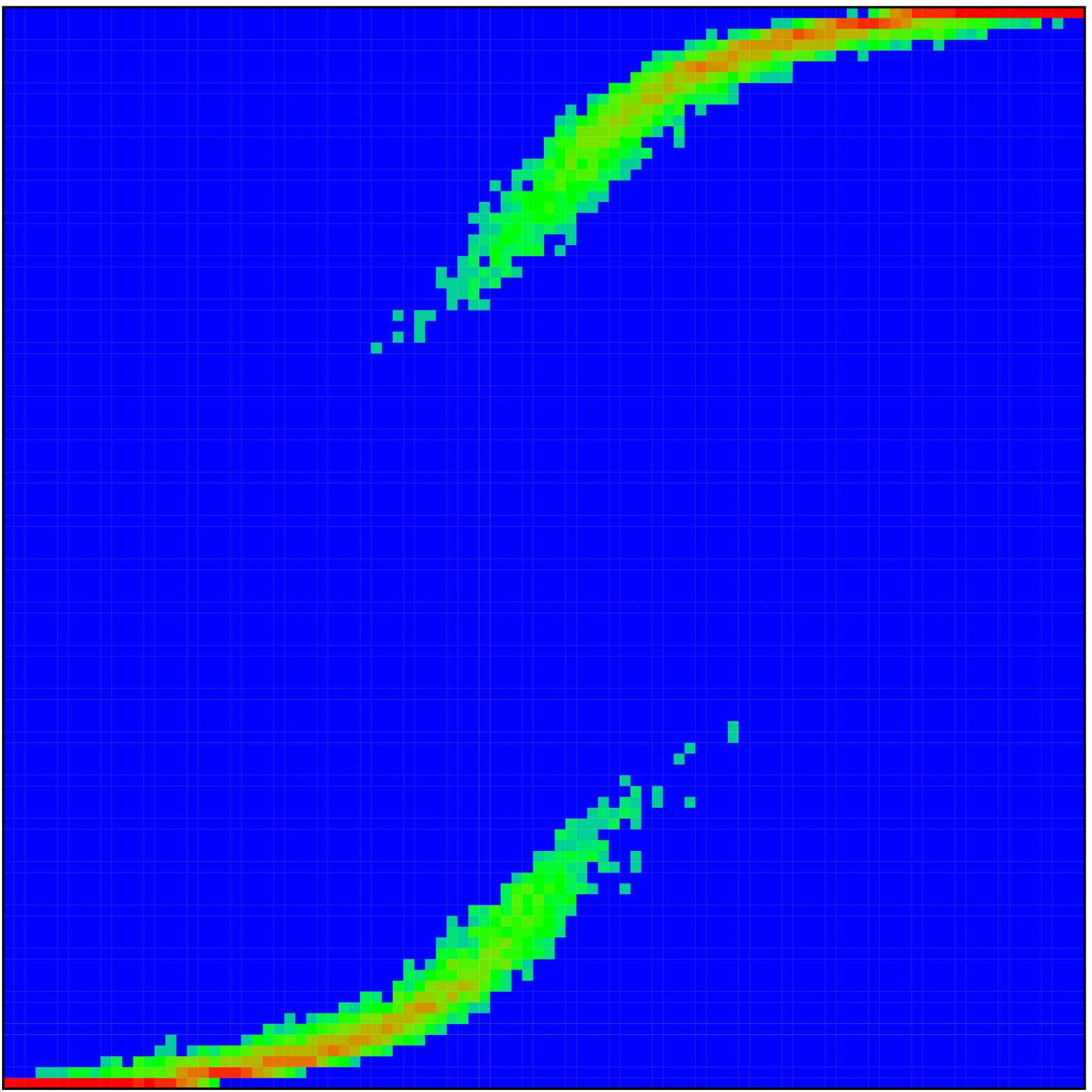}
\end{center}
\vglue -0.2cm
\caption{(Color online) Same as in Fig.~\ref{fig10} 
but for PROF-Sznajd model, option 2.
\label{fig11}} 
\end{figure}

The dependence of the final fraction of red nodes $f_f$ on 
its initial value $f_i$ is shown by the density plot of
probability $W_f$ in Fig.~\ref{fig10} (option 1 of PROF-Sznajd model).
The probability $W_f$ is obtained from many initial random realizations
in a similar way to the case of Fig.~\ref{fig3}. We see that 
there is a significant difference compared to the PROF model (Fig.~\ref{fig3}):
now even at small values of $f_i$ we find small but finite
values of $f_f$ while in the PROF model the red color disappears at 
$f_i < f_c$. This feature is related to the essence of the Sznajd model:
here even small groups can resist against totalitar opinion.
Other features of Fig.~\ref{fig10} are similar to 
those found for the PROF model:
we again observe bistability of opinion formation. The number of nodes $N_g$,
which form the group, does not affect significantly the distribution
$W_f$, we have smaller fluctuations at larger $N_g$ values
but the model works in a stable way already at $N_g=3$.
The results for the option 2 of PROF-Sznajd model are shown in Fig.~\ref{fig11}.
In this case the opinions with a small initial fraction of red nodes $f_i$
are suppressed in a significantly stronger way compared to the option 1.
We attribute this to the fact that large groups can suppress in a stronger way
small groups since the outgoing direct links are taken into 
account in this option.

\begin{figure}[ht]
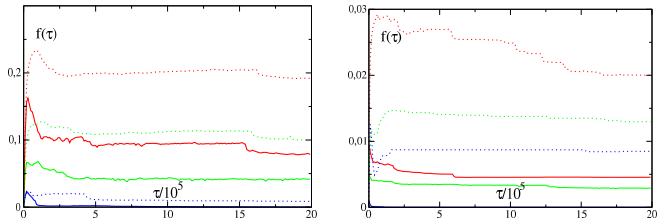

\vglue +0.2cm
\begin{center}
\includegraphics[width=0.23\textwidth]{fig12a}
{\hskip 0.2cm}
\includegraphics[width=0.23\textwidth]{fig12b}
\end{center}
\vglue -0.2cm
\caption{(Color online) Time evolution of the fraction of red nodes
$f(\tau)$ in the PROF-Sznajd model with the initial
red nodes for the top  PageRank nodes:
$N_{top}=200$ (blue/black);
$1000$ (green/light gray);
$2000$ (red/gray); here $N_g=8$.
Full/dashed curves are for Cambridge/Oxford networks;
left panel is for option 1; right panel is for option 2. 
Color of curves is red, green, blue from top to bottom
at maximal $\tau$ on both panels.
\label{fig12}}
\end{figure}

The significant difference between the two options of PROF-Sznajd model
is well seen from the data of Fig.~\ref{fig12}. Here, all $N_{top}$
nodes are taken in red (compare with the PROF model in Fig.~\ref{fig4}).
For the option 1 the society elite succeeds to impose its opinion to a 
significant fraction of nodes which is increased by a factor 5-10.
Visibly, this increase is less significant than in the PROF model.
However, for the option 2 of PROF-Sznajd model
there is practically no increase of the fraction of red nodes.
Thus, in the option 2 the society members are very independent and
the influence of the elite on their opinion is very weak.

\section{V. PROF models on the LiveJournal network}

Even if one can expect that the properties of University networks
are similar to those of the real social networks 
it is important to analyze the previous PROF models in the frame of real
social network. For that we use
the LiveJournal network collected, described and presented at \cite{benczur}.
From this database we obtain the directed network
with $N=3577166$ nodes, $N_{\ell}=44913072$ links which are mainly directed
(only about 30\% of links are symmetric).
The Google matrix of the network is constructed in a usual way \cite{meyerbook}
and its PageRank vector is determined by the iteration process
at the damping factor $\alpha=0.85$.  
For the time evolution of fraction of red nodes $f$ 
we use time iterations in $t$ and $\tau$  defined 
as in previous Sections.
\begin{figure}[ht]
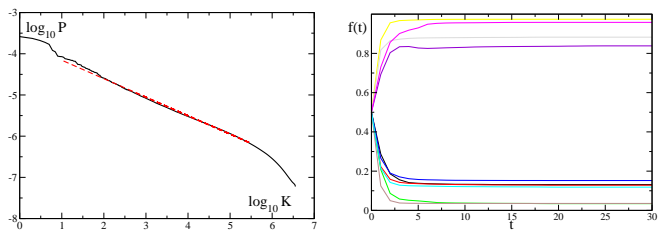

\vglue +0.2cm
\begin{center}
\includegraphics[width=0.23\textwidth]{fig13a}
{\hskip 0.2cm}
\includegraphics[width=0.23\textwidth]{fig13b}
\end{center}
\vglue -0.2cm
\caption{(Color online) Data for LiveJournal network.
{\it Left panel:} PageRank probability
decay with PageRank index $K$ (full curve),
the fitted algebraic dependence is shown by the dashed line 
$y=b-\beta x$ (for $1 \leq \log_{10}K \leq 5.5$)
with the
exponent $\beta=0.448 \pm 0.000046$ and $b=-3.70 \pm 0.00023$.
{\it Right panel:} Time evolution of opinion given
 by a fraction of  red nodes $f(t)$
as a function of number of iterations $t$ (cf. Fig.~\ref{fig1})
at $a=0.5$,  few random initial realizations 
with $f_i=0.5$ are shown.
\label{fig13}}
\end{figure}

The PageRank probability decay $P(K)$ is shown in Fig.~\ref{fig13}.
It is well described by an algebraic law $P(K) \propto 1/K^{\beta}$
with $\beta = 0.448 \pm 0.000046$. The convergence of 
a fraction of red nodes $f(t)$
takes place approximately on the same convergence time 
scale $t_c \sim 5 \sim O(1)$ even if the size of the networks is 
increased almost by a factor 20.

\begin{figure}[ht]
\vglue +0.2cm
\begin{center}
\includegraphics[width=0.23\textwidth]{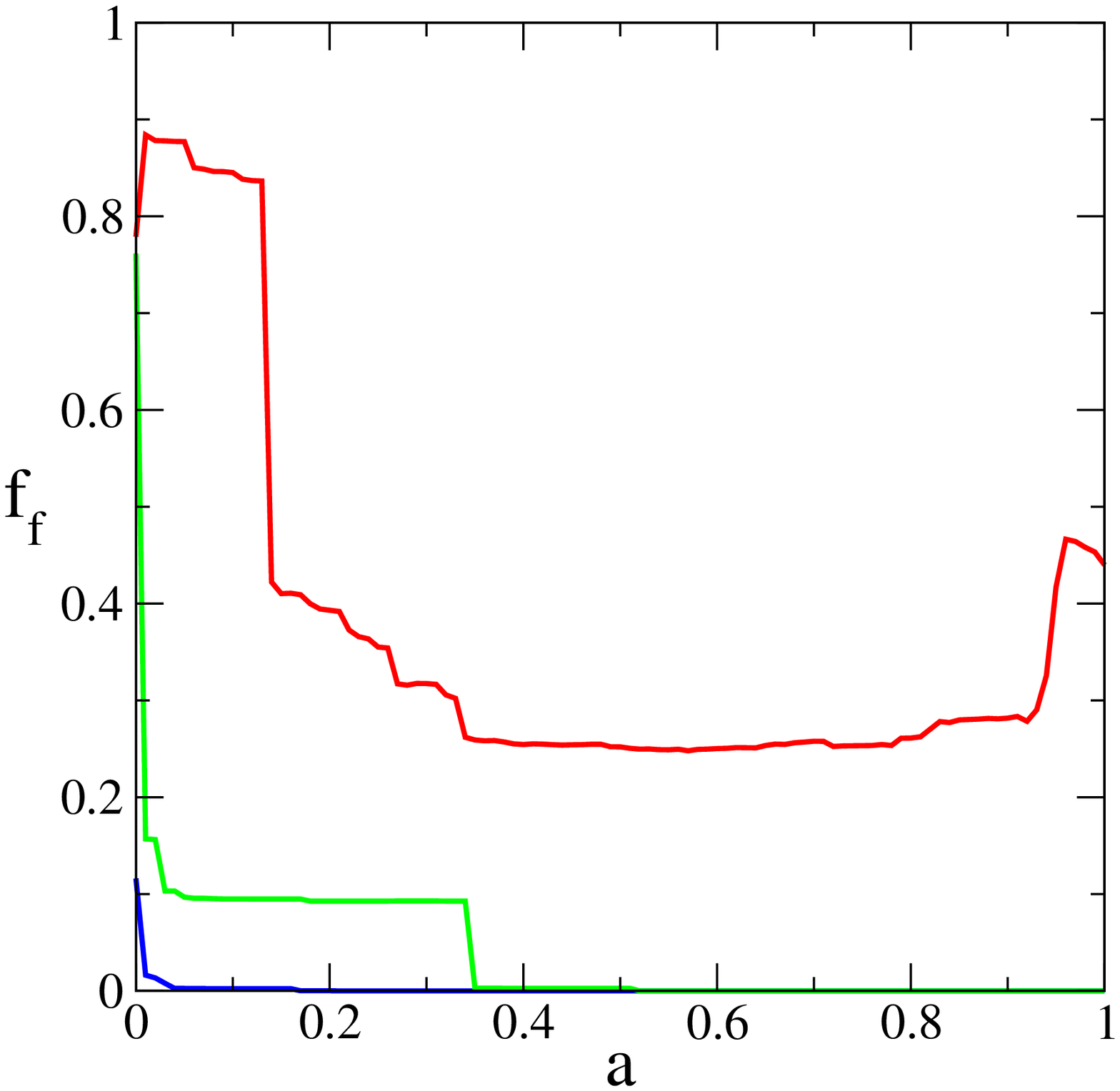}
\includegraphics[width=0.23\textwidth]{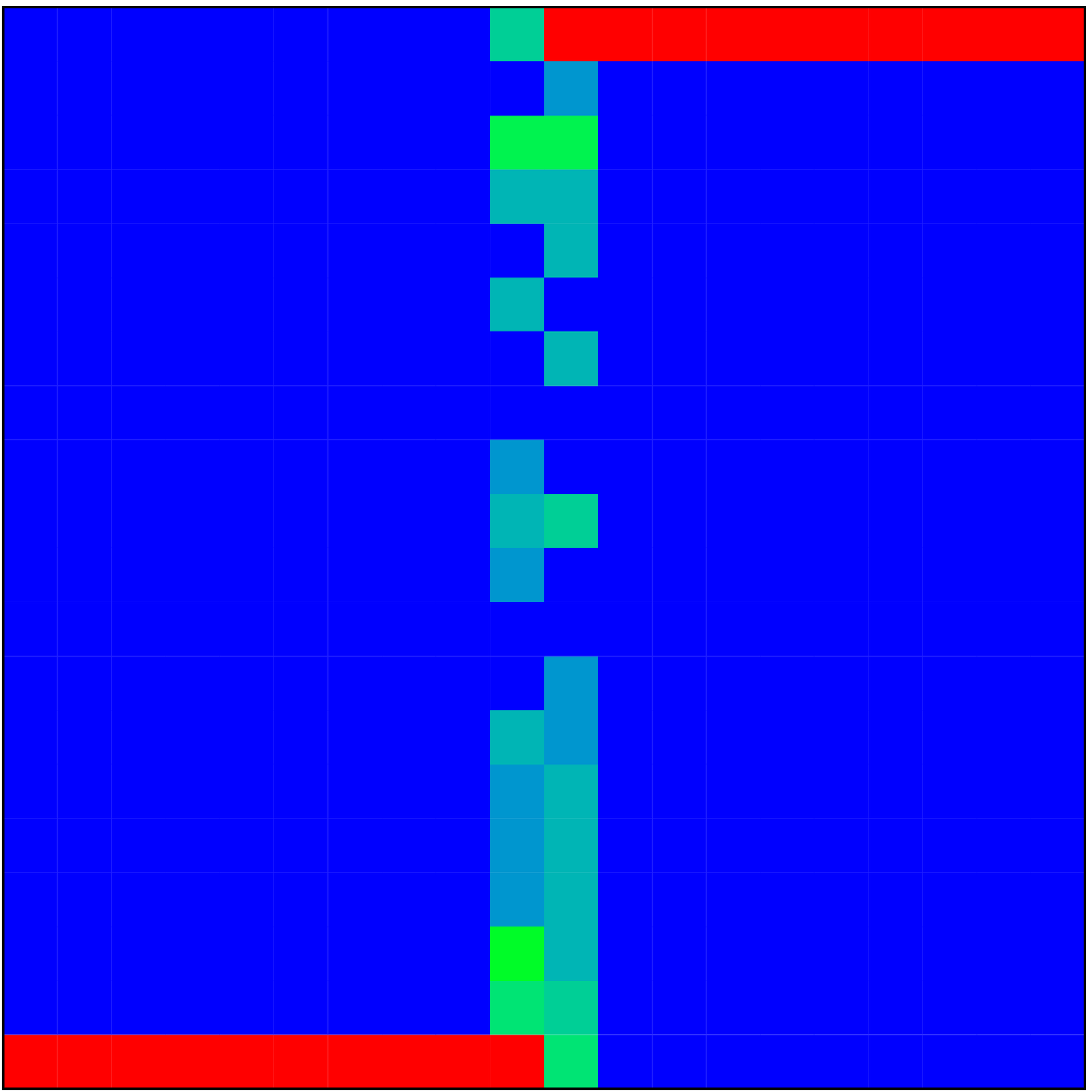}
\end{center}
\vglue -0.2cm
\caption{(Color online) Data for LiveJournal network. {\it Left panel:}
Dependence of the final fraction of red nodes $f_f$
on the tenacious parameter $a$ (or conformist parameter $b=1-a$)
in the PROF model
for initial red nodes in  $N_{top}$ values of PageRank index
($1\leq K \leq N_{top}$, cf. Fig.~\ref{fig4}).
Here, $N_{top}=2000$ blue; $10000$ green,
$35000$  red curves (from bottom to top at $a=0.5$);
$T=0$. {\it Right panel:} Same data as in Fig.~\ref{fig3}
at $a=0.5$ with same parameters but for LiveJournal network.
\label{fig14}}
\end{figure}

In a way similar to the University networks we find that
the homogeneous opinion of society elite 
presented in a small fraction of $N_{top}$ nodes
influences a large fraction of the whole society
especially when the parameter $a$ is not very large
(see Fig.~\ref{fig14} in comparison with Fig.~\ref{fig4}).
The influence of the elite at 1\% of red nodes is 
larger in the case of LiveJournal network.
It is possible that this is related to a 
30\% larger number of links but it is also possible that
other structural network parameters also play
a role here.

\begin{figure}[ht]
\vglue +0.2cm
\begin{center}
\includegraphics[width=0.23\textwidth]{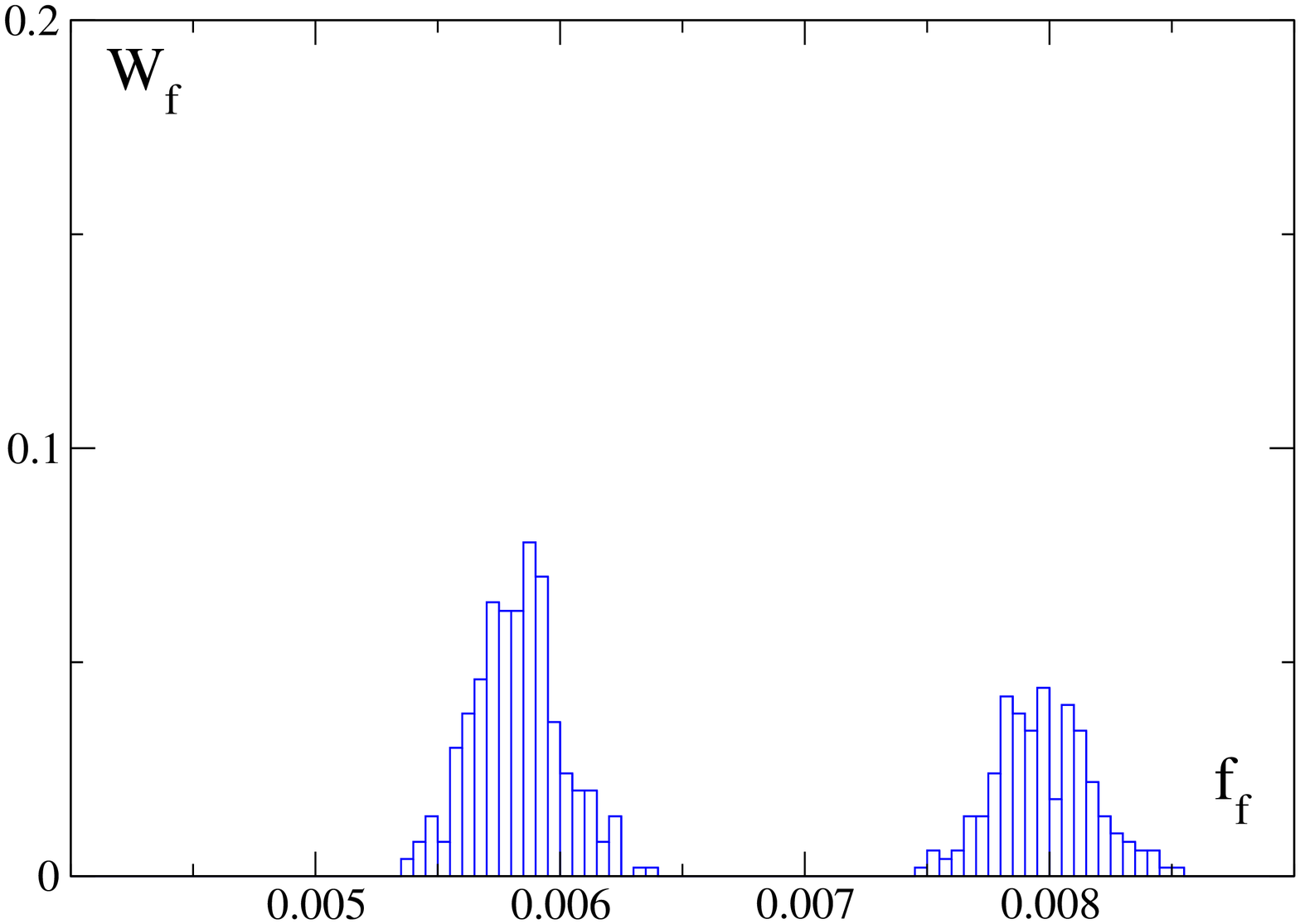}
\hskip 0.3cm
\includegraphics[width=0.23\textwidth]{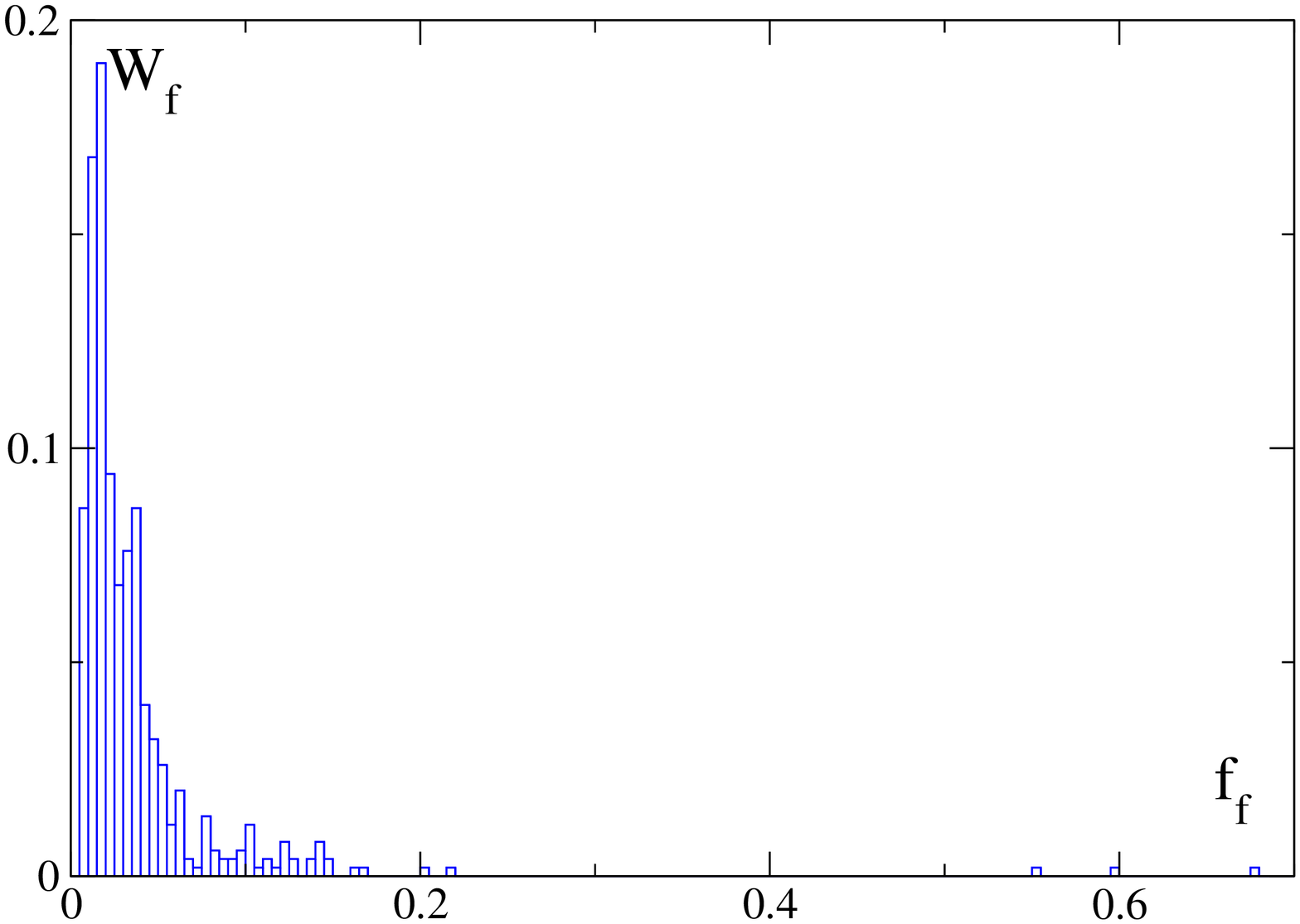}
\end{center}
\vglue -0.2cm
\caption{(Color online) Data for LiveJournal network,
probability distribution $W_f$ of final opinion $f_f$
for a fixed initial opinion $f_i=0.4$ and $a=0.5$ in the PROF model.
{\it Left panel:} usual linear weight $P(K)$
in Eq.(\ref{eq1}). {\it Right panel:} a quadratic weight $P^2(K)$
in Eq.(\ref{eq1}). Histograms are obtained 
with $N_r=500$ initial random realizations,
the normalization is fixed by the condition that
the sum of $W_f$ over all histogram bins is equal to unity.
\label{fig15}}
\end{figure}

In spite of certain similarities with the previous data
for university networks discussed before we find that the
opinion diagram for the LiveJournal network
(see Fig.~\ref{fig14} right panel) is very different from those
obtained for the University networks (see Fig.~\ref{fig3}):
the bistability practically disappeared.
We think that this difference originates from a 
significantly slower decay exponent for PageRank $P(K)$
in the case of   LiveJournal. To check this assumption we
compare the probability distribution $W_f$
of final opinion $f_f$ for an initial opinion 
fixed at $f_i=0.4$ using the PROF model with
the usual linear weight $P$ in Eq.(\ref{eq1})
and a quadratic weight proportional to $P^2$ (see Fig.~\ref{fig15}).
For the linear weight we find that only very small
values of $f_f \approx 0.005$ can be found for initial $f_i=0.4$
while for the quadratic weight we obtain a rather 
broad distribution of $f_f$
values in the main range $0 < f_f < 0.15$
with a few large values $f_f \approx 0.6$.
Thus we see that the final opinion is rather sensitive
to the weight used in Eq.(\ref{eq1}). 
However, in contrast to the University networks
(see Fig.~\ref{fig3},Fig.~\ref{fig5}), where we
have narrow one peak or double peak  distributions of $f_f$,
for the LiveJournal network with quadratic weight we find
a rather broad distribution of $f_f$. In the spirit
of a renormalization map description considered in \cite{galamepl}
(see Figs.1,2 there),
it is possible to assume that one or two peaks corresponds to one or two
fixed points attractor of the map. We make a conjecture that
a broad distribution as in Fig.~\ref{fig15} (right panel)
can correspond to a regime of strange chaotic attractor
appearing in the renormalization map dynamics. In principle, such
a chaotic renormalization dynamics is known to appear in coupled spins 
lattices when   three-spin couplings are present
(see \cite{sargis} and Refs. therein). It is possible that 
a presence of weight probability associated with the PageRank in a certain
power may lead to a chaotic dynamics which would generate a broad 
distribution of final opinions $f_f$.

We also made tests for the PROF-Sznajd model (option 1)
for the LiveJournal database. However, in this case
at $f_i=0.4$, $a=0.5$ we found only small $f_f$ values
(similar of those in Fig.~\ref{fig15}, left panel)
both for linear and quadratic weights in Eq.(\ref{eq1}).
It is possible that the Sznajd groups are less sensitive
to the probability weight.

\section{VI. PROF models for  the Twitter dataset}
We also analyzed the opinion formation on the
Twitter dataset with $N= 41652230$, $N_\ell= 1468365182$
taken from \cite{twitters}. This size is rather large
and due to that we present only main features
of the PROF model for this directed network.
\begin{figure}[ht]
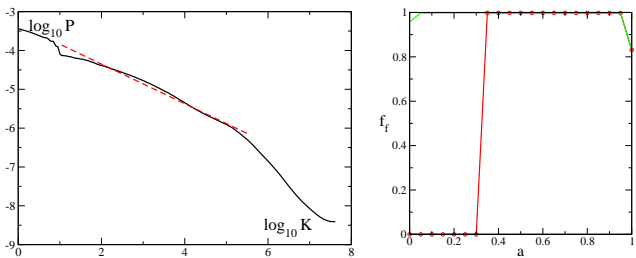

\vglue +0.2cm
\begin{center}
\includegraphics[width=0.26\textwidth]{fig16a}
{\hskip 0.1cm}
\includegraphics[width=0.19\textwidth]{fig16b}
\end{center}
\vglue -0.2cm
\caption{(Color online) Data for Twitter network.
{\it Left panel:} PageRank probability
decay with PageRank index $K$ (full curve),
the fitted algebraic dependence is shown by the dashed line 
$y=b-\beta x$ (for $1 \leq \log_{10}K \leq 5.5$)
with the
exponent $\beta=0.511 \pm 0.0021$ and $b=-3.33 \pm 0.0069$
(for the range $5.5 \leq \log_{10}K \leq 7$ we find
$\beta=1.23$).
{\it Right panel:} Dependence of the final fraction of red nodes $f_f$
on the tenacious parameter $a$ (or conformist parameter $b=1-a$)
in the PROF model
for initial red nodes in  $N_{top}$ values of PageRank index
($1\leq K \leq N_{top}$, cf. Fig.~\ref{fig4}, Fig.~\ref{fig14}).
Here, $N_{top}=1200$ (blue line at $f_f=0$); $1250$ (red curve with circles),
$1300$ (top gree line);
$T=0$.
\label{fig16}}
\end{figure}

The dependence of PageRank $P$ on its index $K$ is shown in Fig.~\ref{fig16} (left panel).
For the range $1 \leq \log_{10} K \leq 5.5$ we find the decay exponent
$\beta \approx 0.51$, being similar to those of LiveJournal network (see Fig.~\ref{fig13})
even if there is a faster drop of $P$ at larger $K$ values.
We note that the value $\beta \approx0.5$ is rather different from the value
usually found for the Zipf law \cite{zipf} and the WWW \cite{meyerbook}
with $\beta \approx 1$. It is possible that this is related to a significantly 
larger average number of links per node which is increased by a factor 3.5 for
the Twitter network compared to the University networks analyzed in 
the previous Sections.

The effect of the homogeneous elite opinion 
of all red $N_{top}$ nodes is shown in Fig.~\ref{fig16} (right panel).
We see that on the Twitter network a small fraction of elite
with the fixed opinion 
($N_{top}/N \approx 3 \times 10^{-5}$)
can impose it practically to the whole community
for all values of the conformist parameter $1-a$.
We find that for $N_{top} > 1300$ all $f_f$ values are 
very close to unity,
while for $N_{top} < 1200$ we find $f_f=0$ 
as it is seen in Fig.~\ref{fig16}, right panel.
Thus, the transition is very sharp.
We attribute such a strong influence of elite opinion
to the very connected structure of Twitter network
with a significantly larger average number of links
per node comparing to the University and LiveJournal networks. 

At $a=0.5$, for a fixed fraction of initial opinion $f_i=0.4$,
we find that the probability distribution $W_f$
of final opinion $f_f$ is located in the range
of small values $0.0006 < W_f <0.0007$ both for
linear $P$ and quadratic $P^2$ weight used in Eq.(\ref{eq1})
(we do not show these data).
For the linear weight the situation is rather similar to the case
of LiveJournal (see Fig.~\ref{fig15}), but for the quadratic weight
we find a significant difference between two networks  (see Fig.~\ref{fig15}).
The reason of such a significant difference 
for the quadratic weight case requires a more detailed comparison
of network properties.

The large size of Twitter network
makes numerical simulations of the PROF-Sznajd model
rather heavy and due to that we did not study
this model for this network.

\section{VII. Discussion}

In this work we proposed the PageRank model of opinion formation
of social networks and analyzed its properties on example of four
different networks. For two University networks
we find rather similar properties of opinion formation.
It is characterized by the important feature according to which
the society elite  with a fixed opinion
can impose it to a significant fraction of the society members
which is much larger than the initial elite fraction.
However, when the initial opinions of society members,
including the elite,
are presented by two options
then we find a significant range of opinion fraction
within a bistability regime. This range
depends on the conformist parameter which
characterizes the local aspects of opinion
formation of linked society members.
The generalization of the Sznajd model
for the scale-free social networks 
gives  interesting examples of opinion formation
where finite small size groups can keep their
own opinion being different from the main opinion of the majority.
In this way the proposed PROF-Sznajd model 
shows that the totalitar opinions can be escaped by small
sub-communities.
We find that the properties of opinion formation 
are rather similar for the two University networks
of Cambridge and Oxford.
However, the results obtained for networks
of LiveJournal and Twitter show that the range
of bistability practically disappears for these networks.
Our data indicate that this is related to a slower algebraic
decay of PageRank in these cases compared to the University
networks. However, the deep reasons of such a difference
require a more detailed analysis. Indeed,
LiveJournal and Twitter networks 
demonstrate rather different behavior 
for the $P^2$-weighted function of opinion formation.
The studies performed for regular networks \cite{galamepl}  
show existence of stable or bistable fixed points
for opinion formation models
that have certain similarities with the opinion formation
properties found in our studies. At the same time
the results obtained in \cite{sargis} show that three-body spin coupling
can generate a chaotic renormalization dynamics.
Some our results (Fig.~\ref{fig15}, right panel)
give indications on a possible existence of such 
chaotic phase on the social networks.

The enormous development of social networks
in a few last years \cite{livejournal,facebook,twitter,vkontakte}
definitely shows that the analysis of opinion formation
on such networks requires further investigations.
This research can find also  various other applications.
One of them can be a neuronal network of brain
which represents itself a directed scale-free network \cite{apkarian}.
The applications of network science  to brain networks
is now under a rapid development  (see e.g. \cite{olaf})
and the Google matrix methods can find useful
applications in this field \cite{gbrain}.

This work is supported in part by the EC FET Open project 
``New tools and algorithms for directed network analysis''
(NADINE $No$ 288956). We thank A.Bencz\'ur and S.Vigna for 
providing us a friendly access
to the LiveJournal database \cite{benczur}
and the Twitter dataset \cite{twitters}.

\end{document}